\documentclass[runningheads]{llncs}

\usepackage[mobile]{eccv}
\usepackage{eccvabbrv}
\usepackage{graphicx}
\usepackage{booktabs}
\usepackage{graphicx}
\usepackage{multirow}
\usepackage{makecell}
\usepackage{colortbl}
\usepackage{amsmath}
\usepackage{amssymb}
\usepackage{mathtools}
\usepackage[accsupp]{axessibility}
\usepackage{hyperref}
\usepackage{orcidlink}

\begin{document}
\title{Federated Learning for Blind Image Super-Resolution} 
\titlerunning{Federated Learning for Blind Image Super-Resolution}
\author{\underline{Brian B. Moser}\inst{1, 2, 3}\orcidlink{0000-0002-0290-7904} \and
\underline{Ahmed Anwar}\inst{1, 3}\orcidlink{0009-0004-9737-7177} \and
Federico Raue\inst{1}\orcidlink{0000-0002-8604-6207}\and
Stanislav Frolov\inst{1,2}\orcidlink{0000-0003-2700-5031}\and
Andreas Dengel\inst{1,2}\orcidlink{0000-0002-6100-8255}
}

\authorrunning{Moser et al.}
\institute{
German Research Center for Artificial Intelligence (DFKI), Germany \and
RPTU Kaiserslautern-Landau, Germany \and
Equal Contribution\\
\email{first.second@dfki.de}}

\maketitle

\begin{abstract}
Traditional blind image SR methods need to model real-world degradations precisely.
Consequently, current research struggles with this dilemma by assuming idealized degradations, which leads to limited applicability to actual user data.
Moreover, the ideal scenario - training models on data from the targeted user base - presents significant privacy concerns.
To address both challenges, we propose to fuse image SR with federated learning, allowing real-world degradations to be directly learned from users without invading their privacy.
Furthermore, it enables optimization across many devices without data centralization. 
As this fusion is underexplored, we introduce new benchmarks specifically designed to evaluate new SR methods in this federated setting. 
By doing so, we employ known degradation modeling techniques from SR research. 
However, rather than aiming to mirror real degradations, our benchmarks use these degradation models to simulate the variety of degradations found across clients within a distributed user base. 
This distinction is crucial as it circumvents the need to precisely model real-world degradations, which limits contemporary blind image SR research.
Our proposed benchmarks investigate blind image SR under new aspects, namely differently distributed degradation types among users and varying user numbers.
We believe new methods tested within these benchmarks will perform more similarly in an application, as the simulated scenario addresses the variety while federated learning enables the training on actual degradations.
\end{abstract}

\section{Introduction}
Billions of photos are taken daily, representing a rich and untapped resource.
The potential to use these images as training samples to enhance Super-Resolution (SR) techniques is immense.
Image SR is the task of generating High-Resolution (HR) images from their Low-Resolution (LR) counterparts and is a critical component in applications like medical imaging, remote sensing, and consumer electronics \cite{10041995,moser2024diffusion,el2023single,wang2022review}.
Yet, the complex nature of real-world image degradation - comprising blur, noise, and compression artifacts - poses a significant challenge to image SR.

In the literature, researchers work with simulated degradations that might be encountered in practical applications. 
However, accurately predicting real-world degradations from HR to LR images remains challenging and depends on the user base. 
For example, some studies assume JPEG compression \cite{liu2020learning,kong2022reflash}, while others do not \cite{gu2019blind,zhang2018learning}. 
Ultimately, the relevance of JPEG compression in training depends on the specific application and the characteristics of the data of the respective user base \cite{liu2022blind}.
As SR is meant to be applied by users, the optimal strategy would be to train on the given user data, bypassing the need to model the degradation encountered.
To safely utilize actual user data, we propose the adoption of Federated Learning (FL), which ensures privacy and data security \cite{federatedlearning17,fl2024survey}.
It enables the collaborative training of SR models across multiple devices while keeping the data localized.
Overall, the fusion of image SR and FL brings the following benefits: 
\begin{itemize}
    \setlength\itemsep{0em}
    \item Direct and local training of SR models on end-user data;
    \item Global model training without exposing user-sensitive data by using FL;
    \item Improved robustness to various degradations not part of the training data by aggregating model weights from different clients with different degradations;
\end{itemize}

Unfortunately, the fusion of FL with image SR is limited \cite{FeSR_LIDAR} and, to the best of our knowledge, non-existent under the lens of unknown degradations across different users.
The goal of FL in image SR is to collaboratively train a global model that is robust to various degradations.
To bridge the gap of missing work in this setting, we introduce new benchmarks designed to evaluate emerging SR methods in this specific FL setting.
These benchmarks aim to empirically assess the influence of the number of participating clients on model performance and to explore the effects of varying degradation patterns across user data. 
We present preliminary assessments of SR models, namely SRResNet \cite{ledig2017photo} and RRDB \cite{wang2018esrgan}, which will serve as a foundation for further research and opportunities for improvement in this field.
Through this approach, we seek to shed light on the dynamics of federated SR systems and their adaptability to real-world degradation scenarios.
Our initial experiments showed that more clients lead to robust global models against complex degradation.
We also observe that the distribution of degradations is important: noisy and JPEG-compressed data proves more critical than blurred and clean images.
Lastly, we propose an evaluation in a one-client setting to identify an idealistic (centralized) upper-bound performance.

\section{Related Work}
\subsection{Blind Image Super-Resolution}
Traditional SR approaches often assume a specific degradation, i.e., bicubic downsampling, which fails to address complex degradations encountered in real-world images \cite{10041995,moser2024diffusion,liu2022blind}. 
This limitation has inspired blind SR research that addresses unknown degradations.
Based on the degradation modeling, methods can be roughly categorized into \textbf{explicit modeling}, \textbf{implicit modeling}, and those utilizing \textbf{internal statistics of single images} \cite{liu2022blind}.

\textbf{Explicit modeling strategies}, such as SRMD \cite{zhang2018learning} and IKC \cite{gu2019blind}, attempt to parameterize the degradation process directly, often requiring external datasets for model training. 
They have to identify and model the specific types of degradation present correctly.
These methods have shown proficiency in addressing a range of degradation types but are constrained by the need for accurate degradation estimation, which makes a real-world application challenging.

\textbf{Implicit modeling approaches}, exemplified by CinCGAN \cite{yuan2018unsupervised} and FSSR \cite{fritsche2019frequency}, leverage external datasets to learn degradation distributions indirectly. 
These methods have demonstrated versatility in handling real-world degradations but are challenged by the intrinsic limitations of GANs, which can introduce artifacts into the SR results \cite{10041995}.

Furthermore, methods capitalizing on the \textbf{internal statistics of single images}, such as ZSSR \cite{shocher2018zero} and KernelGAN \cite{bell2019blind}, offer promising avenues for blind SR by exploiting the redundancy within a given image. 
Thus, they do not require any external training data.
However, they are also inherently limited by the image content, showing variability in performance across different image types and making real-world applications time-consuming because of multiple training epochs to super-resolve a single image.

In summary, developing methods that can robustly handle varying real-world degradations without extensive reliance on external data or suffering from GAN-induced artifacts remains challenging.
We address this by introducing FL, which can exploit real-world data directly without hurting privacy concerns.

\subsection{Federated Learning}
\textit{McMahan et al.} coined the term for Federated Learning (FL) and presented evidence that model averaging, known as FedAvg, provides a useful algorithm to aggregate knowledge distributed among different entities \cite{federatedlearning17}. 
Since then, FL has gone through extensive research, especially for data heterogeneity \cite{fedprox, astraea}, and data privacy \cite{mcmahan2017privacyFL, bhowmick2019privacyFL}.  
On the other hand, fields such as image classification \cite{astraea} and text prediction \cite{mcmahan2017privacyFL} have demonstrated the benefits of FL. The potential of utilizing federated learning in the field of image SR is, however, yet to be explored.
In our work, we mainly leverage the privacy guarantees of FL while shedding light on data heterogeneity.
Data heterogeneity, also known as the Non-IID (independent and identically distributed) problem, has been recognized as the main obstacle for FL applications since it challenges one of the main assumptions of machine learning, which is that data is IID \cite{fl2024survey}. 
FedProx \cite{fedprox} was introduced to deal with this problem by adding an $l_2$ regularization term to penalize local models from deviating too far away from the global model. 
SCAFFOLD \cite{scaffold}  applied stochastic variance reduction to control drift during training, while FedPVR \cite{fedpvr} utilized the same concept on a subset of the layers focusing on removing variance in later feature extraction layers since earlier layers learn more abstract and straightforward features.
While these methods aim to mitigate the statistical diversity of the training data distributions, we aim to exploit this diversity to train image SR models to be more robust against different types of degradations.

\begin{figure}[!t]
  \centering
  \includegraphics[width=\linewidth]{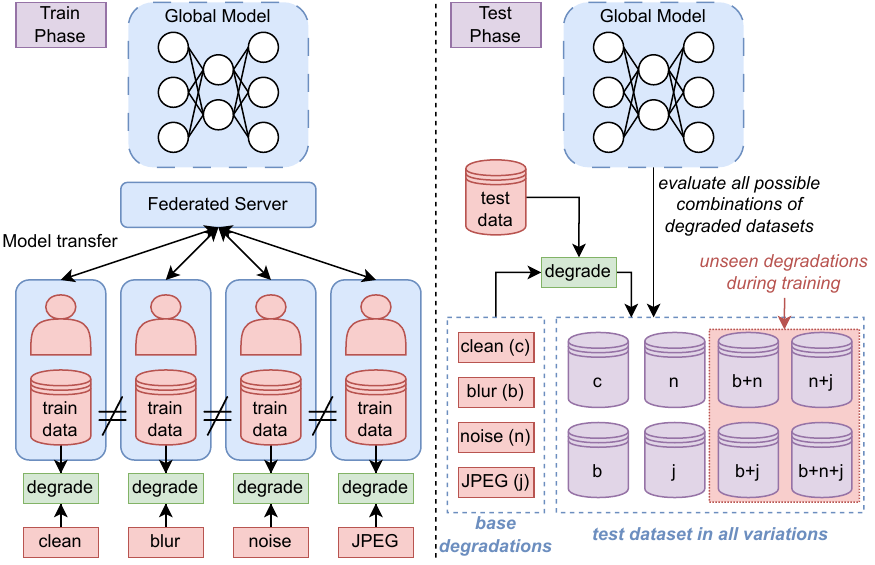}
  \caption{\label{fig:main}Proposed setup of FL for blind image SR. 
  Each client contributes by training on their local data and individual degradation type to the collaborative training of a global model.
  The federated server collects the training results, updates the global model, and synchronizes the global model with the clients.
  During testing, the trained global model is evaluated on variations of the test dataset with all possible combinations of degradations.
}
\end{figure}

\section{Federated Learning for Blind Image SR}
Current blind image SR methods, despite being an improvement over classical SR in mimicking real degradations, still model degradations they might encounter in practice.
Consequently, they yield models optimized for assumed conditions, which may not align with actual data. 
Ideally, the optimal strategy employs training directly on the respective application's user data to ensure maximal quality. 
Yet, this conflicts with the imperative need to uphold user privacy, a significant challenge in harmonizing model performance with data confidentiality.
Our proposed solution is to use FL for training image SR models as it allows the training directly on actual data and guarantees privacy and data safety.
Thus, it facilitates access to a broader range of data that would otherwise remain inaccessible.
Overall, our introduction of this new setting invites fresh perspectives but also comes with challenges that necessitate evaluation.
This section will address the most critical aspects, namely how to train on different degradation types and investigate different dataset degradation distributions.

\subsection{General Setting}
The goal of FL for blind image SR is to train a global model that is robust to various degradations collaboratively.
The federated server saves and synchronizes the global model with its clients, as illustrated in \autoref{fig:main}.
The clients train the current model on their local data in each training round with individual degradations. 
We assume that data is not shared between clients to simulate real-world scenarios.
Clients return the updated weights to the server after training without sharing any information about their data.
The server aggregates the results and updates the global model by averaging them \cite{federatedlearning17}.
The cycle repeats until convergence or a fixed number of training rounds. 

\subsection{Implicit High-Order Degradation Modeling}
\label{sec:hodm}
We adopt a proven approach known as high-order degradation modeling, initially introduced by \textit{Wang et al.}, for handling multiple degradation settings \cite{wang2021real,wang2018basicsr,kong2022reflash}. 
In this setting, one employs complex combinations of various basic degradation types: clean (no degradation), blur, noise, and JPEG compression.
During training, the kernels, downsampling scales, noise levels, and compression rates are typically selected randomly and applied on the fly.
We then derive a training LR image $\mathbf{x}$ from the HR image $\mathbf{y}$ via
\begin{equation}
    \label{equ:JPEG_model}
    \centering
    \mathbf{x}=((\mathbf{y}\otimes \boldsymbol{k})\downarrow_s+\boldsymbol{n})_{JPEG_{q}},
\end{equation} 
where $\boldsymbol{k}$ is a blur kernel, $\boldsymbol{n}$ the additive noise, $s$ the scaling factor, and \emph{q} the quality factor of a JPEG compression.

We, however, exploit \autoref{equ:JPEG_model} to simulate the diversity of possible degradation conditions by assigning each client of our FL setting exactly one degradation type before training.
Thus, each client trains on a single degradation type locally but never on a combination of degradation types, as illustrated on the left side of \autoref{fig:main}.
Letting each client contribute only one degradation type leads to a meaningful evaluation against combinations of degradations that have not been encountered during training (cross-degradation evaluation).

For instance, one client possesses only clean images, while another has only blurry images, and so on.
This differs from the regular, one-client high-order degradation modeling in a non-FL setting in the literature, where multiple degradations are combined during training \cite{liu2022blind}.
Consequently, the global model implicitly learns the high-order degradation modeling by aggregating the results of the degradation-diverse clients.
We will test the robustness of the global model by applying the full combination range of degradations in \autoref{equ:JPEG_model}.
In other words, we will evaluate all degradation combinations during the testing phase, as illustrated on the right side of \autoref{fig:main}.
For example, one variant of a test dataset has only clean images, another one has images that are both blurred and JPEG-compressed, and so on.
We will further evaluate our results against a one-client, centralized setting as upper-bound.

\subsection{Degradation and Data Distribution}
\label{sec:dirichlet}
Sensitive private data, i.e., a highly private image, is usually stored by one user. 
Therefore, we simulate a real-world training scenario by storing one image exclusively in one client.
In practice, however, some users might have more images than others.
Since we also assume that one client has precisely one degradation type to ensure a meaningful cross-degradation evaluation (see \autoref{sec:hodm}), we can not generally assume a uniform distribution.
Some clients have more data, thereby contributing more of their assigned degradation type to the collaborative training. 
We simulate skewed distributions by using the Dirichlet distribution \cite{ContrastiveFL, FedDM}:
\begin{equation}
    \centering
     f(d_1,\ldots,d_K; \alpha_1, \ldots, \alpha_K) = \frac{1}{\text{B}(\alpha)} \prod_{i=1}^{K}\mathbf{x}_i^{\alpha_i - 1}
\end{equation} 
\noindent
where the concentration parameter $\alpha_i$ controls how uniformly the degradation type $d_i$ is distributed. 
If $\alpha_i$ is high, the data is more uniformly distributed. 
A lower $\alpha_i$ means some degradation types will be more common than others. 
We follow the standard procedure in FL by setting $\alpha=0.5$ \cite{ContrastiveFL, FedDM}.
The normalizing factor is a multivariate Beta distribution, which can be expressed in terms of the Gamma function as:

\begin{equation}
    \centering
     \text{B}(\alpha) = \frac{\prod_{i=1}^{K}\Gamma(\alpha_i)}{\Gamma(\Sigma_{i=1}^K \alpha_i)} ,\quad \alpha = (\alpha_i, \ldots, \alpha_K)
\end{equation} 

By employing skewed distributions through the Dirichlet distribution, we can simulate a real-world scenario in which data is not uniformly distributed among users. 
This distribution analysis allows us to understand how varying amounts of data with different types of degradation impact the learning process. 
Typically, we might expect that skewed distributions, where certain degradation types are over-represented, could lead to a biased global model towards these degradations. 
Conversely, underrepresented degradations might not be learned as effectively, potentially reducing the overall robustness and performance on degradation-diverse data as faced in our cross-degradation evaluation, i.e., combinations of degradations. 
This exploration is crucial for developing SR models that perform well across a broad range of degradations encountered by users.

\section{Experiments}
In this section, we present the results of our benchmarks and give hands-on recommendations on how to perform future experiments in this federated setting.
Our benchmarks evaluate varying numbers of clients, different degradation type distributions, and a comparison against an idealized (centralized) setting as the upper bound.
We further provide preliminary assessments of SR models as a foundation for further research.
Additional hyper-parameters and implementation details can be found in the supplementary material.

\textbf{Datasets.}
We used 800 2K resolution images from DIV2K for training \cite{agustsson2017ntire}.
For a fair comparison, we distribute each image to precisely one client, so no two clients have the same training image.
For testing, we used the Set5~\cite{Set5}, Set14~\cite{Set14}, BSD100~\cite{BSD100}, Manga109~\cite{Manga109}, and Urban100~\cite{Urban100} datasets and created all possible degradation combinations of them according to \autoref{equ:JPEG_model}.
We apply $4\times$ upscaling tasks in all experiments.

\textbf{Architectures.}
We evaluate two representative SR networks - SRResNet \cite{ledig2017photo}  with 1.5M parameters and RRDB \cite{wang2018esrgan} with 16.7M parameters.
Our findings can broadly apply to other CNN-based SR networks that feature similar architectural designs \cite{wang2021real,zhang2021designing}.

\subsection{Results}

\begin{figure}[!t]
  \centering
  \includegraphics[width=\linewidth]{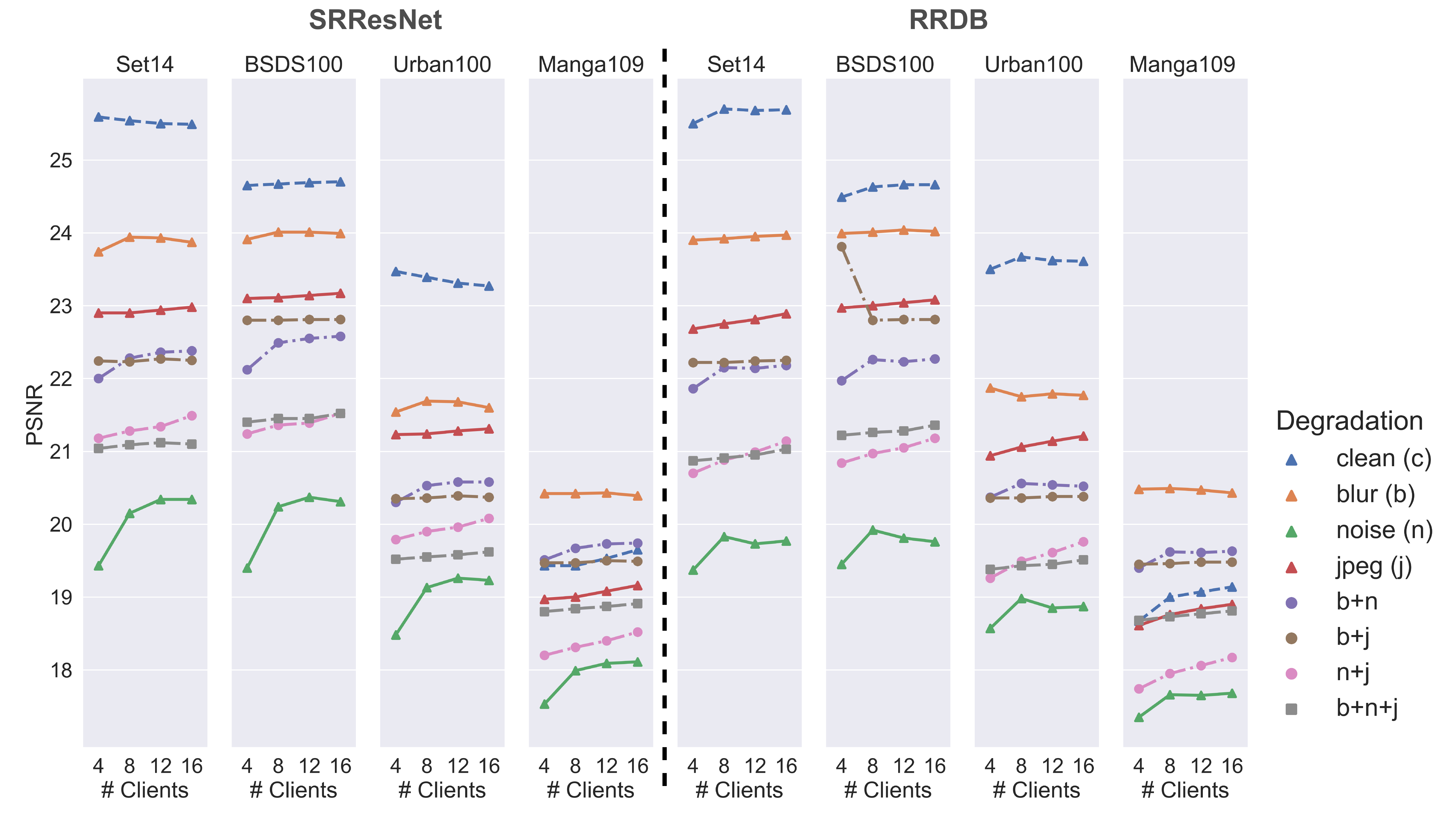}
  \vskip -0.3cm
  \caption{\label{fig:degradations_srresnet}The PSNR (db) results of SRResNet and RRDB with $\times 4$ scaling and varying number of clients (4, 8, 12, 16) on different test datasets (Set14, BSD100, Urban100, Manga109).
  We tested the SR model on eight degradation variations listed in the legend on the right side (clean, blur, noise, JPEG, and combinations). The degradation legend shows triangles, circles, and squares for single, double, and triple degradations, respectively.
}
\end{figure}

\textbf{Different Client Numbers.}
\autoref{fig:degradations_srresnet} evaluates the influence of different numbers of clients with SRResNet and RRDB. 
Note that the data distribution, i.e., which HR image undergoes which degradation, remains the same (uniform distribution) across all experiments in this section.
For SRResNet, as the number of clients increases, variation in PSNR values is observed across all datasets.
Interestingly, the model robustness improves for combined degradations, such as those using all degradations simultaneously.
This starkly contrasts the diminished PSNR values noted for images with clean or singularly blurred degradations, which worsen as more clients participate in the model training.
In parallel assessments using the RRDB model, a similar theme of variability emerges with an expanded pool of clients, albeit with some distinctions. 
Specifically, the model demonstrates a decline in handling noise-induced degradations and exhibits varied results for images degraded by blur and noise. 
Despite these challenges, the overall trend towards enhanced robustness against composite degradations persists, echoing the observations made with SRResNet.

\textit{Benchmark-Recommendation:} The results on both SR models indicate that the number of clients has a regularizing effect on complex degradations (two and three degradations), even though the data distribution, i.e., which HR image undergoes which degradation, remains the same across all experiments.
We recommend reproducing this experiment for new methods, especially those introducing new regularization. 

\begin{figure}[t!]
  \centering
  \includegraphics[width=\linewidth]{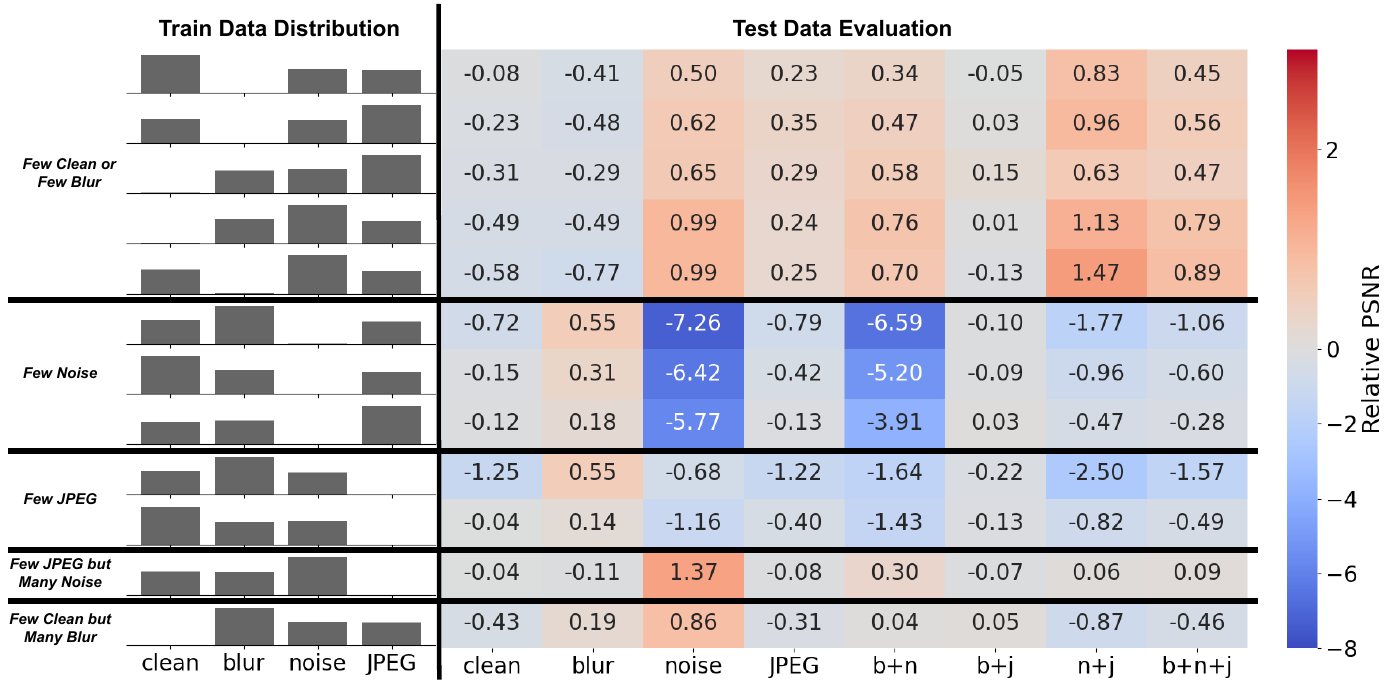}
  \vskip -0.3cm
  \caption{\label{fig:dataDistribution}Experimental results of different degradation distributions on BSD100. 
  On the left side is the degradation distribution during training.
  The resulting performance on the respective degraded test set is on the right side (relative to uniform distribution).
  We identified five clusters that exhibit similar performance variations (see left side: Few Clean or Few Blur; Few Noise; Few JPEG; Few JPEG but Many Noise, Few Clean but Many Blur). 
}
\end{figure}

\textbf{Different Degradation Distributions.}
We analyze the impact of varied degradation type distributions on BSD100 using the Dirichlet distribution (see \autoref{sec:dirichlet}).
This procedure generates 24 distinct train dataset distributions.
The skewed distributions of the four degradation types emphasize one degradation while suppressing another. 
The remaining two degradations maintain a similar level.
Given the minimal variation between two similar-level degradation types, we streamlined the analysis by focusing only on half of the train dataset distributions. 
This decision was based on the observation that similar degradation distributions yield comparable results.
We encountered five patterns, via agglomerative clustering of the test set results, for the remaining distributions, as depicted in \autoref{fig:dataDistribution}. 
The first cluster contains either a low amount of clean or blur images. 
Accordingly, the performance drops slightly on clean or blurred images and improves on all other complex combinations.
Likewise, another cluster, which has nearly no clean images but many blurry images, leads to performances similar to a uniform distribution.
This highlights that the amount of clean or blurry images does not have the same impact as noisy or JPEG images.
Another cluster, which either has a low amount of noise or JPEG images, provides further evidence for this observation.
It leads to a decline in performance in almost all combinations.
We can conclude that noise and JPEG degradations are crucial for model robustness.
With the last cluster, we can observe that more noisy images can compensate for the absence of JPEG images.

\textit{Benchmark-Recommendation:} We recommend focusing on one training distribution per cluster to reduce the number of experiments in the future from 24 to 5: Few Clean or Few Blur; Few Noise; Few JPEG; Few JPEG but Many Noise; Few Clean but Many Blur.

\begin{table*}[t!]
  \caption{SRResNet comparison between a one-client versus FL setting (with 16 clients). 
  }
  \small
  \begin{center}
    \begin{tabular}{|c|cc|cc|cc|cc|}
      \hline
        & \multicolumn{2}{c|}{Set14~\cite{Set14}} & \multicolumn{2}{c|}{BSD100~\cite{BSD100}} & \multicolumn{2}{c|}{Urban100~\cite{Urban100}} & \multicolumn{2}{c|}{Manga109~\cite{Manga109}} \\ \hline
      Clients & \multicolumn{1}{c|}{clean} & blur & \multicolumn{1}{c|}{clean} & blur & \multicolumn{1}{c|}{clean} & blur & \multicolumn{1}{c|}{clean} & blur \\ \hline
      1 & \multicolumn{1}{c|}{25.97} & 25.93 & \multicolumn{1}{c|}{24.86} & 24.90 & \multicolumn{1}{c|}{23.99} & 23.71 & \multicolumn{1}{c|}{19.42} & 19.55 \\ 
      16 & \multicolumn{1}{c|}{25.49} & 23.87 & \multicolumn{1}{c|}{24.70} & 23.99 & \multicolumn{1}{c|}{23.27} & 21.60 & \multicolumn{1}{c|}{19.65} & 20.39 \\\hline \hline 
      difference &  \multicolumn{1}{c|}{-0.48} & -2.06 & \multicolumn{1}{c|}{-0.16} & \multicolumn{1}{c|}{-0.72} & \multicolumn{1}{c|}{-2.11} & -0.91 & \multicolumn{1}{c|}{+0.23} & +0.84 \\\hline \hline
      
      Clients &  \multicolumn{1}{c|}{noise} & JPEG & \multicolumn{1}{c|}{noise} & JPEG & \multicolumn{1}{c|}{noise} & JPEG & \multicolumn{1}{c|}{noise} & JPEG \\ \hline
      1 &  \multicolumn{1}{c|}{22.45} & 23.45 & \multicolumn{1}{c|}{22.53} & 23.45 & \multicolumn{1}{c|}{21.03} & 21.73  & \multicolumn{1}{c|}{19.29} & 19.29\\
      16 &  \multicolumn{1}{c|}{20.34} & 22.98 & \multicolumn{1}{c|}{20.31} & 23.17  & \multicolumn{1}{c|}{19.23} & 21.31 & \multicolumn{1}{c|}{18.11} & 19.16 \\ \hline \hline
      difference & \multicolumn{1}{c|}{-2.11} & -0.47 & \multicolumn{1}{c|}{-2.22} & -0.28  & \multicolumn{1}{c|}{-1.80} & -0.42 & \multicolumn{1}{c|}{-1.18} & -0.13 \\\hline \hline
      
      Clients  & \multicolumn{1}{c|}{b+n} & b+j & \multicolumn{1}{c|}{b+n} & b+j & \multicolumn{1}{c|}{b+n} & b+j & \multicolumn{1}{c|}{b+n} & b+j \\ \hline
      1 &  \multicolumn{1}{c|}{21.92} & 22.53 & \multicolumn{1}{c|}{22.43} & 22.98 & \multicolumn{1}{c|}{20.30} & 20.58 & \multicolumn{1}{c|}{19.46} & 19.76 \\ 
      16 &  \multicolumn{1}{c|}{22.38} & 22.25 & \multicolumn{1}{c|}{22.58} & 22.81  & \multicolumn{1}{c|}{20.58} & 20.37 & \multicolumn{1}{c|}{19.74} & 19.49 \\ \hline \hline
      difference &  \multicolumn{1}{c|}{+0.46} & -0.28 & \multicolumn{1}{c|}{+0.15} & -0.17 & \multicolumn{1}{c|}{+0.28} & -0.21 & \multicolumn{1}{c|}{+0.28} & -0.27 \\\hline \hline
      
      Clients  & \multicolumn{1}{c|}{n+j} & b+n+j & \multicolumn{1}{c|}{n+j} & b+n+j & \multicolumn{1}{c|}{n+j} & b+n+j & \multicolumn{1}{c|}{n+j} & b+n+j \\ \hline
      1 &  \multicolumn{1}{c|}{22.07} & 21.65 & \multicolumn{1}{c|}{22.02} & 21.92  & \multicolumn{1}{c|}{20.61} & 19.96 & \multicolumn{1}{c|}{18.75} & 19.25 \\
      16 &  \multicolumn{1}{c|}{21.49} & 21.10 & \multicolumn{1}{c|}{21.52} & 21.52  & \multicolumn{1}{c|}{20.08} & 19.62 & \multicolumn{1}{c|}{18.52} & 18.91 \\\hline\hline 
      difference &  \multicolumn{1}{c|}{-0.58} & -0.55 & \multicolumn{1}{c|}{-0.50} & -0.40 & \multicolumn{1}{c|}{-0.53} & -0.34 & \multicolumn{1}{c|}{-0.23} & -0.34 \\\hline 
      \end{tabular}
  \end{center}
  \vskip -0.3cm
  
  \label{table:degradations_centralized_srresnet}
  \vskip -0.49cm
\end{table*}
\begin{figure}[!t]
  \centering
  \includegraphics[width=.89\linewidth]{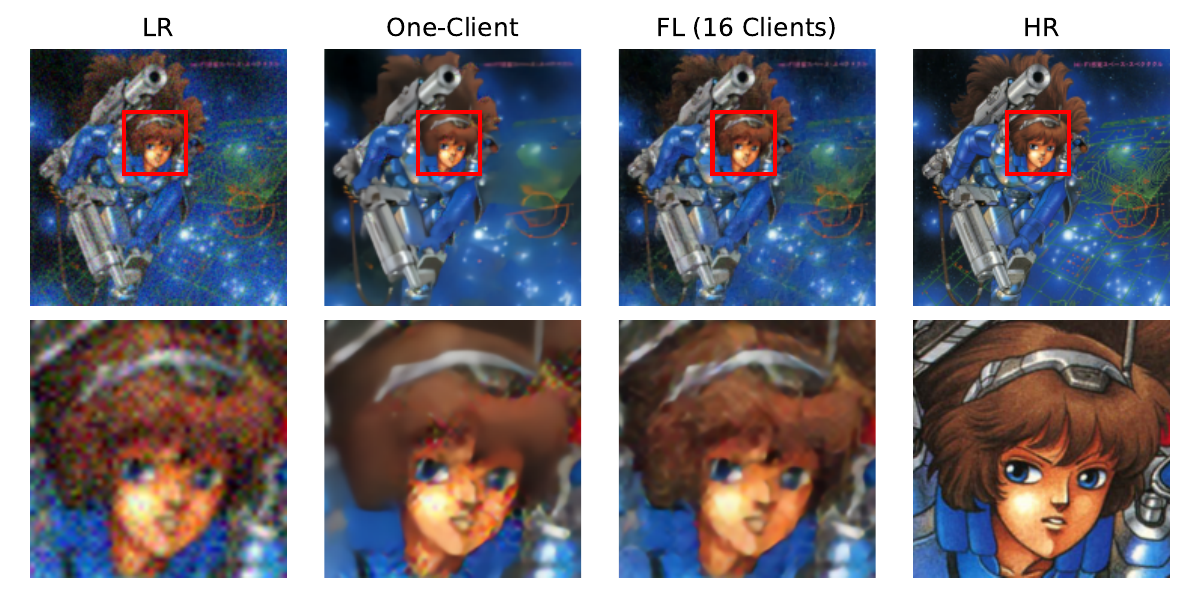}
  \vskip -0.3cm
  \caption{\label{fig:manga}Visual results on Manga109 with blur and noise degradations in the LR image.
}
\end{figure}

\textbf{Centralized Scenario.}
\autoref{table:degradations_centralized_srresnet} contrasts FL against the idealized scenario of a one-client (centralized) setting, which serves as an upper bound due to its access to all user data at the cost of privacy. 
The PSNR for noise and blurred images is consistently lower than the centralized setting across all datasets. 
Conversely, the difference is less pronounced for clean and JPEG degradations, indicating that FL training is relatively more resilient to these degradations. 
This suggests that training data across multiple clients introduces challenges in maintaining performance on less complex degradations, which aligns with the observations made while evaluating varying numbers of clients.
Another common observation with the varying numbers of clients is that the model demonstrates an improvement or minimal decrease in performance for combined degradations (b+n, b+j, n+j, and b+n+j) across most datasets. 
This improvement suggests that FL, with a larger and more diverse client base, can potentially enhance its robustness to complex real-world degradations.
Interestingly, the Manga109 dataset exhibited unique behavior with improved PSNR for blur degradation. 
This dataset, known for its high-contrast and detailed illustrations, might benefit from the diversified degradation types introduced by FL. 
\autoref{fig:manga} illustrates this finding.

\textit{Benchmark-Recommendation:} We recommend this experiment as it shows a potential upper bound performance of the federated setting. Moreover, we suggest investigating Manga109 more intensively in the future within this setting.

\section{Limitations}
One significant constraint is the extensive VRAM requirements of the simulations.
SRResNet needs approximately 20GB per client.
This requirement becomes more demanding with RRDB, requiring ca. 40GB.
Such high VRAM demands present a challenge, particularly for researchers with limited access to high-end hardware.
Future work should explore more memory-efficient simulation techniques that reduce VRAM usage without compromising the quality of the FL setting. 
Real-world applications are not affected by this limitation.
\section{Conclusion}
Current research is limited by assuming idealized degradations, hindering applicability to actual user data.
To bridge this gap, we propose to fuse FL with blind image SR, an uncharted territory that presents a potential avenue for novel ideas and perspectives.
This fusion circumvents the traditional necessity for precise degradation modeling by directly learning from real-world degradations in user data while maintaining user privacy.
We present comprehensive benchmarks and recommendations for future work tailored to explore blind image SR within an FL context, emphasizing handling diverse degradation types and the participation of varied user numbers.
Our findings reveal that SR models trained with FL show potential for robust performance across a spectrum of degradations, suggesting a promising direction for future SR applications that demand adaptability to user-specific degradation patterns. 
By evaluating the impact of client numbers and degradation distributions, we aim to provide a foundation for future research that can further refine the potential of FL for blind image SR.
Overall, FL with blind image SR opens new pathways for research and application, promising enhanced model robustness, user privacy, and, ultimately, a closer alignment of SR methods with the complexities of real-world image degradation.

\section*{Acknowledgment}
This work was supported by the BMBF project SustainML (Grant 101070408) and the BMWK project EuroDaT (Grant 68GX21010K).

\bibliographystyle{splncs04}
\bibliography{main}

\begin{thebibliography}{10}
\providecommand{\url}[1]{\texttt{#1}}
\providecommand{\urlprefix}{URL }
\providecommand{\doi}[1]{https://doi.org/#1}

\bibitem{agustsson2017ntire}
Agustsson, E., Timofte, R.: Ntire 2017 challenge on single image super-resolution: Dataset and study. In: CVPRW. pp. 126--135 (2017)

\bibitem{bell2019blind}
Bell-Kligler, S., Shocher, A., Irani, M.: Blind super-resolution kernel estimation using an internal-gan. NeurIPS  \textbf{32} (2019)

\bibitem{beutel2020flower}
Beutel, D.J., Topal, T., Mathur, A., Qiu, X., Fernandez-Marques, J., Gao, Y., Sani, L., Kwing, H.L., Parcollet, T., Gusmão, P.P.d., Lane, N.D.: Flower: A friendly federated learning research framework. arXiv preprint arXiv:2007.14390  (2020)

\bibitem{Set5}
Bevilacqua, M., Roumy, A., Guillemot, C., Alberi-Morel, M.L.: Low-complexity single-image super-resolution based on nonnegative neighbor embedding  (2012)

\bibitem{bhowmick2019privacyFL}
Bhowmick, A., Duchi, J., Freudiger, J., Kapoor, G., Rogers, R.: Protection against reconstruction and its applications in private federated learning (2019)

\bibitem{astraea}
Duan, M.: Astraea: Self-balancing federated learning for improving classification accuracy of mobile deep learning applications. CoRR  \textbf{abs/1907.01132} (2019), \url{http://arxiv.org/abs/1907.01132}

\bibitem{el2023single}
El-Shafai, W., Ali, A.M., El-Nabi, S.A., El-Rabaie, E.S.M., Abd El-Samie, F.E.: Single image super-resolution approaches in medical images based-deep learning: a survey. Multimedia Tools and Applications pp. 1--37 (2023)

\bibitem{fritsche2019frequency}
Fritsche, M., Gu, S., Timofte, R.: Frequency separation for real-world super-resolution. In: 2019 IEEE/CVF International Conference on Computer Vision Workshop (ICCVW). pp. 3599--3608. IEEE (2019)

\bibitem{FeSR_LIDAR}
Gkillas, A., Arvanitis, G., Lalos, A.S., Moustakas, K.: Federated learning for lidar super resolution on automotive scenes. In: 2023 24th International Conference on Digital Signal Processing (DSP). pp.~1--5 (2023). \doi{10.1109/DSP58604.2023.10167942}

\bibitem{gu2019blind}
Gu, J., Lu, H., Zuo, W., Dong, C.: Blind super-resolution with iterative kernel correction. In: CVPR. pp. 1604--1613 (2019)

\bibitem{Urban100}
Huang, J.B., Singh, A., Ahuja, N.: Single image super-resolution from transformed self-exemplars. In: CVPR. pp. 5197--5206 (2015)

\bibitem{scaffold}
Karimireddy, S.P., Kale, S., Mohri, M., Reddi, S.J., Stich, S.U., Suresh, A.T.: {SCAFFOLD:} stochastic controlled averaging for on-device federated learning. CoRR  \textbf{abs/1910.06378} (2019), \url{http://arxiv.org/abs/1910.06378}

\bibitem{kong2022reflash}
Kong, X., Liu, X., Gu, J., Qiao, Y., Dong, C.: Reflash dropout in image super-resolution. In: CVPR. pp. 6002--6012 (2022)

\bibitem{ledig2017photo}
Ledig, C., Theis, L., Husz{\'a}r, F., Caballero, J., Cunningham, A., Acosta, A., Aitken, A., Tejani, A., Totz, J., Wang, Z., et~al.: Photo-realistic single image super-resolution using a generative adversarial network. In: CVPR. pp. 4681--4690 (2017)

\bibitem{fedpvr}
Li, B., Schmidt, M.N., Alstr{\o}m, T.S., Stich, S.U.: Partial variance reduction improves non-convex federated learning on heterogeneous data. ArXiv  \textbf{abs/2212.02191} (2022), \url{https://api.semanticscholar.org/CorpusID:254246803}

\bibitem{ContrastiveFL}
Li, Q., He, B., Song, D.: Model-contrastive federated learning. In: CVPR. pp. 10713--10722 (June 2021)

\bibitem{liu2022blind}
Liu, A., Liu, Y., Gu, J., Qiao, Y., Dong, C.: Blind image super-resolution: A survey and beyond. IEEE TPAMI  \textbf{45}(5),  5461--5480 (2022)

\bibitem{liu2020learning}
Liu, P., Zhang, H., Cao, Y., Liu, S., Ren, D., Zuo, W.: Learning cascaded convolutional networks for blind single image super-resolution. Neurocomputing  \textbf{417},  371--383 (2020)

\bibitem{BSD100}
Martin, D., Fowlkes, C., Tal, D., Malik, J.: A database of human segmented natural images and its application to evaluating segmentation algorithms and measuring ecological statistics. In: ICCV. vol.~2, pp. 416--423. IEEE (2001)

\bibitem{Manga109}
Matsui, Y., Ito, K., Aramaki, Y., Fujimoto, A., Ogawa, T., Yamasaki, T., Aizawa, K.: Sketch-based manga retrieval using manga109 dataset. Multimedia Tools and Applications  \textbf{76}(20),  21811--21838 (2017)

\bibitem{federatedlearning17}
McMahan, H.B., Moore, E., Ramage, D., y~Arcas, B.A.: Federated learning of deep networks using model averaging. CoRR  \textbf{abs/1602.05629} (2016), \url{http://arxiv.org/abs/1602.05629}

\bibitem{mcmahan2017privacyFL}
McMahan, H.B., Ramage, D., Talwar, K., Zhang, L.: Learning differentially private language models without losing accuracy. CoRR  \textbf{abs/1710.06963} (2017), \url{http://arxiv.org/abs/1710.06963}

\bibitem{10041995}
Moser, B.B., Raue, F., Frolov, S., Palacio, S., Hees, J., Dengel, A.: Hitchhiker's guide to super-resolution: Introduction and recent advances. IEEE TPAMI  \textbf{45}(8),  9862--9882 (Aug 2023)

\bibitem{moser2024diffusion}
Moser, B.B., Shanbhag, A.S., Raue, F., Frolov, S., Palacio, S., Dengel, A.: Diffusion models, image super-resolution and everything: A survey. arXiv preprint arXiv:2401.00736  (2024)

\bibitem{fedprox}
Sahu, A.K., Li, T., Sanjabi, M., Zaheer, M., Talwalkar, A., Smith, V.: On the convergence of federated optimization in heterogeneous networks. CoRR  \textbf{abs/1812.06127} (2018), \url{http://arxiv.org/abs/1812.06127}

\bibitem{fl2024survey}
Shao, J., Li, Z., Sun, W., Zhou, T., Sun, Y., Liu, L., Lin, Z., Mao, Y., Zhang, J.: A survey of what to share in federated learning: Perspectives on model utility, privacy leakage, and communication efficiency (2024)

\bibitem{shocher2018zero}
Shocher, A., Cohen, N., Irani, M.: “zero-shot” super-resolution using deep internal learning. In: CVPR. pp. 3118--3126 (2018)

\bibitem{wang2021real}
Wang, X., Xie, L., Dong, C., Shan, Y.: Real-esrgan: Training real-world blind super-resolution with pure synthetic data. In: ICCV. pp. 1905--1914 (2021)

\bibitem{wang2018basicsr}
Wang, X., Yu, K., Chan, K.C., Dong, C., Loy, C.C.: Basicsr: Open source image and video restoration toolbox. GitHub: San Francisco, CA, USA  (2018)

\bibitem{wang2018esrgan}
Wang, X., Yu, K., Wu, S., Gu, J., Liu, Y., Dong, C., Qiao, Y., Change~Loy, C.: Esrgan: Enhanced super-resolution generative adversarial networks. In: Proceedings of the European conference on computer vision (ECCV) workshops. pp.~0--0 (2018)

\bibitem{wang2022review}
Wang, X., Yi, J., Guo, J., Song, Y., Lyu, J., Xu, J., Yan, W., Zhao, J., Cai, Q., Min, H.: A review of image super-resolution approaches based on deep learning and applications in remote sensing. Remote Sensing  \textbf{14}(21), ~5423 (2022)

\bibitem{FedDM}
Xiong, Y., Wang, R., Cheng, M., Yu, F., Hsieh, C.J.: Feddm: Iterative distribution matching for communication-efficient federated learning. In: CVPR. pp. 16323--16332 (June 2023)

\bibitem{yuan2018unsupervised}
Yuan, Y., Liu, S., Zhang, J., Zhang, Y., Dong, C., Lin, L.: Unsupervised image super-resolution using cycle-in-cycle generative adversarial networks. In: CVPRW. pp. 701--710 (2018)

\bibitem{Set14}
Zeyde, R., Elad, M., Protter, M.: On single image scale-up using sparse-representations. In: International conference on curves and surfaces. pp. 711--730. Springer (2010)

\bibitem{zhang2021designing}
Zhang, K., Liang, J., Van~Gool, L., Timofte, R.: Designing a practical degradation model for deep blind image super-resolution. In: ICCV. pp. 4791--4800 (2021)

\bibitem{zhang2018learning}
Zhang, K., Zuo, W., Zhang, L.: Learning a single convolutional super-resolution network for multiple degradations. In: CVPR. pp. 3262--3271 (2018)

\end{thebibliography}

\newpage
\section*{Supplementary Material}

\subsection*{Implementation Details}
The code for our experiments can be found on GitHub\footnote{\url{https://github.com/WILL-BE-IN-FINAL}}.
The details for blind image SR were extracted from \textit{Kong et al.} \footnote{\url{https://github.com/XPixelGroup/RDSR}}, which follows the setup of \textit{Wang et al.}, a standard setting for blind SR \cite{wang2021real,wang2018basicsr,kong2022reflash}.
For the federated learning aspects of our work, we used the framework Flower\footnote{\url{https://github.com/adap/flower}} by \textit{Beutel et al.} \cite{beutel2020flower}.
All experiments were performed on multiple NVIDIA A100-80GB GPUs, some required 8 GPUs in parallel, i.e., evaluations with 16 clients.

\subsection*{Data Preparation}
We follow the standard procedure and extracted $480\times480$ HR crops from DIV2K, and $120\times120$ LR crops while using a stride of 240 and 60, respectively, to extract the $4\times$ upscaling datasets \cite{moser2024diffusion}.
Note that the images were distributed among clients before cropping.
This pre-processing allows faster loading than the full images with 2K resolution during training.
Moreover, we pre-generated the degraded datasets to enhance training speed further.
During the experiments, we extract $128\times128$ HR patches and their corresponding LR patches for training, following \textit{Kong et al.} \cite{kong2022reflash}.
For testing data, no patch extraction was applied, and the full image resolution was used.

\subsection*{Hyper-Parameters for Federated Learning}
Locally and for both network architectures (SRResNet and RRDB), each client trained with a batch size of 16. 
With each training round, we performed training on all clients.
The local epoch amount was set to 1 within each round, whereas the number of rounds was set to 200 for all experiments.
For optimization, we used the ADAM optimizer with default settings and a learning rate of 0.0002.

\subsection*{One Client Evaluation}
In our study, we assessed the performance of a one-client scenario as depicted in \autoref{table:oneClient}, observing that a mixed degradation distribution yields superior PSNR scores than those achieved in a federated learning context. 
Conversely, models trained on a singular type of degradation underperformed in comparison. 
It is important to note that surpassing the one-client setting, commonly used in blind image SR research, is not the primary goal of our study. 
Instead, we aim to pioneer the integration of federated learning into the blind image SR domain. 
Although potentially less efficient, this approach introduces significant advantages, including secure data sharing and utilizing client-specific degradations, thus prioritizing privacy and data diversity.

\begin{table*}[t!]
  \caption{The PSNR (db) results of SRResNet with $\times 4$ scaling and varying degradated datasets for one client setting.
  We tested the SR model on test sets with eight different degradation variations (clean, blur, noise, jpeg, and combinations).
  Worse performances compared to 16 clients with SRResNet and uniform degradation distribution are marked in red (see main paper).
  }
  \small
  \begin{center}
  \resizebox{\textwidth}{!}{%
    \begin{tabular}{|c|cc|cc|cc|cc|cc|}
      \hline
       & \multicolumn{2}{c|}{Set5~\cite{Set5}} & \multicolumn{2}{c|}{Set14~\cite{Set14}} & \multicolumn{2}{c|}{BSD100~\cite{BSD100}} & \multicolumn{2}{c|}{Manga109~\cite{Manga109}} & \multicolumn{2}{c|}{Urban100~\cite{Urban100}} \\ \hline
      Dataset & \multicolumn{1}{c|}{clean} & blur & \multicolumn{1}{c|}{clean} & blur & \multicolumn{1}{c|}{clean} & blur & \multicolumn{1}{c|}{clean} & blur & \multicolumn{1}{c|}{clean} & blur \\ \hline
      
      Mix & \multicolumn{1}{c|}{29.68} & 29.50 & \multicolumn{1}{c|}{25.97} & 25.93 & \multicolumn{1}{c|}{24.86} & 24.90 & \multicolumn{1}{c|}{19.42} & 19.55 & \multicolumn{1}{c|}{23.99} & 23.71 \\
      Clean & \multicolumn{1}{c|}{30.04} & 25.97 & \multicolumn{1}{c|}{26.42} & \cellcolor{red!20}23.81 & \multicolumn{1}{c|}{24.98} & 24.03 & \multicolumn{1}{c|}{\cellcolor{red!20}19.37} & 20.50 & \multicolumn{1}{c|}{24.53} & \cellcolor{red!20}21.44 \\
      Blur & \multicolumn{1}{c|}{\cellcolor{red!20}18.47} & 29.89 & \multicolumn{1}{c|}{\cellcolor{red!20}16.99} & 26.23 & \multicolumn{1}{c|}{\cellcolor{red!20}16.99} & 25.03 & \multicolumn{1}{c|}{\cellcolor{red!20}15.31} & \cellcolor{red!20}19.42 & \multicolumn{1}{c|}{\cellcolor{red!20}15.20} & 24.33 \\
      Noise & \multicolumn{1}{c|}{\cellcolor{red!20}24.12} & \cellcolor{red!20}22.52 & \multicolumn{1}{c|}{\cellcolor{red!20}22.78} & \cellcolor{red!20}21.63 & \multicolumn{1}{c|}{\cellcolor{red!20}22.62} & \cellcolor{red!20}21.97 & \multicolumn{1}{c|}{\cellcolor{red!20}19.39} & \cellcolor{red!20}19.57 & \multicolumn{1}{c|}{\cellcolor{red!20}21.44} & \cellcolor{red!20}20.01 \\
      JPEG & \multicolumn{1}{c|}{\cellcolor{red!20}25.91} & \cellcolor{red!20}24.57 & \multicolumn{1}{c|}{\cellcolor{red!20}24.19} & \cellcolor{red!20}23.09 & \multicolumn{1}{c|}{\cellcolor{red!20}23.97} & \cellcolor{red!20}23.40 & \multicolumn{1}{c|}{\cellcolor{red!20}19.14} & \cellcolor{red!20}19.98 & \multicolumn{1}{c|}{\cellcolor{red!20}22.23} & \cellcolor{red!20}21.07 \\ \hline \hline
      
      Dataset & \multicolumn{1}{c|}{noise} & jpeg & \multicolumn{1}{c|}{noise} & jpeg & \multicolumn{1}{c|}{noise} & jpeg & \multicolumn{1}{c|}{noise} & jpeg & \multicolumn{1}{c|}{noise} & jpeg \\ \hline
  
      Mix & \multicolumn{1}{c|}{23.66} & 25.13 & \multicolumn{1}{c|}{22.45} & 23.45 & \multicolumn{1}{c|}{22.53} & 23.45 & \multicolumn{1}{c|}{19.29} & 19.29 & \multicolumn{1}{c|}{21.03} & 21.73 \\
      Clean & \multicolumn{1}{c|}{\cellcolor{red!20}13.74} & \cellcolor{red!20}23.54 & \multicolumn{1}{c|}{\cellcolor{red!20}13.22} & \cellcolor{red!20}22.32 & \multicolumn{1}{c|}{\cellcolor{red!20}13.07} & \cellcolor{red!20}22.77 & \multicolumn{1}{c|}{\cellcolor{red!20}13.33} & \cellcolor{red!20}18.66 & \multicolumn{1}{c|}{\cellcolor{red!20}12.96} & \cellcolor{red!20}20.51 \\
      Blur & \multicolumn{1}{c|}{\cellcolor{red!20}10.64} & \cellcolor{red!20}17.71 & \multicolumn{1}{c|}{\cellcolor{red!20}10.27} & \cellcolor{red!20}17.00 & \multicolumn{1}{c|}{\cellcolor{red!20}10.06} & \cellcolor{red!20}17.71 & \multicolumn{1}{c|}{\cellcolor{red!20}10.47} & \cellcolor{red!20}15.08 & \multicolumn{1}{c|}{\cellcolor{red!20}9.99} & \cellcolor{red!20}14.84 \\
      Noise & \multicolumn{1}{c|}{23.84} & \cellcolor{red!20}23.17 & \multicolumn{1}{c|}{22.55} & \cellcolor{red!20}22.17 & \multicolumn{1}{c|}{22.59} & \cellcolor{red!20}22.36 & \multicolumn{1}{c|}{19.27} & 19.16 & \multicolumn{1}{c|}{21.18} & \cellcolor{red!20}20.90 \\
      JPEG & \multicolumn{1}{c|}{\cellcolor{red!20}16.10} & 25.25 & \multicolumn{1}{c|}{\cellcolor{red!20}15.20} & 23.56 & \multicolumn{1}{c|}{\cellcolor{red!20}14.87} & 23.53 & \multicolumn{1}{c|}{\cellcolor{red!20}15.52} & 19.41 & \multicolumn{1}{c|}{\cellcolor{red!20}14.91} & 21.84 \\ \hline \hline
      
      Dataset & \multicolumn{1}{c|}{b+n} & b+j & \multicolumn{1}{c|}{b+n} & b+j & \multicolumn{1}{c|}{b+n} & b+j & \multicolumn{1}{c|}{b+n} & b+j & \multicolumn{1}{c|}{b+n} & b+j \\ \hline
      
      Mix & \multicolumn{1}{c|}{\cellcolor{red!20}23.21} & 23.86 & \multicolumn{1}{c|}{\cellcolor{red!20}21.92} & 22.53 & \multicolumn{1}{c|}{\cellcolor{red!20}22.43} & 22.98 & \multicolumn{1}{c|}{\cellcolor{red!20}19.46} & 19.76 & \multicolumn{1}{c|}{\cellcolor{red!20}20.30} & 20.58 \\
      Clean & \multicolumn{1}{c|}{\cellcolor{red!20}18.48} & \cellcolor{red!20}23.03 & \multicolumn{1}{c|}{\cellcolor{red!20}17.90} & \cellcolor{red!20}21.92 & \multicolumn{1}{c|}{\cellcolor{red!20}17.85} & \cellcolor{red!20}22.61 & \multicolumn{1}{c|}{\cellcolor{red!20}16.98} & \cellcolor{red!20}19.21 & \multicolumn{1}{c|}{\cellcolor{red!20}17.00} & \cellcolor{red!20}20.05 \\
      Blur & \multicolumn{1}{c|}{\cellcolor{red!20}12.46} & \cellcolor{red!20}20.45 & \multicolumn{1}{c|}{\cellcolor{red!20}11.88} & \cellcolor{red!20}20.11 & \multicolumn{1}{c|}{\cellcolor{red!20}11.78} & \cellcolor{red!20}21.03 & \multicolumn{1}{c|}{\cellcolor{red!20}12.34} & \cellcolor{red!20}17.34 & \multicolumn{1}{c|}{\cellcolor{red!20}11.72} & \cellcolor{red!20}19.09 \\
      Noise & \multicolumn{1}{c|}{\cellcolor{red!20}22.59} & \cellcolor{red!20}21.97 & \multicolumn{1}{c|}{\cellcolor{red!20}21.68} & \cellcolor{red!20}21.23 & \multicolumn{1}{c|}{\cellcolor{red!20}22.02} & \cellcolor{red!20}21.78 & \multicolumn{1}{c|}{\cellcolor{red!20}19.61} & \cellcolor{red!20}19.12 & \multicolumn{1}{c|}{\cellcolor{red!20}20.06} & \cellcolor{red!20}19.64 \\
      JPEG & \multicolumn{1}{c|}{\cellcolor{red!20}20.94} & 23.93 & \multicolumn{1}{c|}{\cellcolor{red!20}19.97} & 22.57 & \multicolumn{1}{c|}{\cellcolor{red!20}19.58} & 23.02 & \multicolumn{1}{c|}{\cellcolor{red!20}18.25} & 19.85 & \multicolumn{1}{c|}{\cellcolor{red!20}18.54} & 20.64 \\ \hline \hline
      
      Dataset & \multicolumn{1}{c|}{n+j} & b+n+j & \multicolumn{1}{c|}{n+j} & b+n+j & \multicolumn{1}{c|}{n+j} & b+n+j & \multicolumn{1}{c|}{n+j} & b+n+j & \multicolumn{1}{c|}{n+j} & b+n+j \\ \hline

      Mix & \multicolumn{1}{c|}{23.37} & 22.80 & \multicolumn{1}{c|}{22.07} & 21.65 & \multicolumn{1}{c|}{22.02} & 21.92 & \multicolumn{1}{c|}{18.75} & 19.25 & \multicolumn{1}{c|}{20.61} & 19.96 \\
      Clean & \multicolumn{1}{c|}{\cellcolor{red!20}21.13} & \cellcolor{red!20}21.34 & \multicolumn{1}{c|}{\cellcolor{red!20}20.51} & \cellcolor{red!20}20.52 & \multicolumn{1}{c|}{\cellcolor{red!20}20.71} & \cellcolor{red!20}20.96 & \multicolumn{1}{c|}{\cellcolor{red!20}17.74} & \cellcolor{red!20}18.40 & \multicolumn{1}{c|}{\cellcolor{red!20}19.08} & \cellcolor{red!20}19.06 \\
      Blur & \multicolumn{1}{c|}{\cellcolor{red!20}14.06} & \cellcolor{red!20}15.26 & \multicolumn{1}{c|}{\cellcolor{red!20}13.57} & \cellcolor{red!20}14.66 & \multicolumn{1}{c|}{\cellcolor{red!20}13.40} & \cellcolor{red!20}14.73 & \multicolumn{1}{c|}{\cellcolor{red!20}13.44} & \cellcolor{red!20}14.45 & \multicolumn{1}{c|}{\cellcolor{red!20}12.69} & \cellcolor{red!20}13.74 \\
      Noise & \multicolumn{1}{c|}{23.33} & \cellcolor{red!20}22.11 & \multicolumn{1}{c|}{22.37} & 21.32 & \multicolumn{1}{c|}{\cellcolor{red!20}22.47} & 21.87 & \multicolumn{1}{c|}{19.14} & 19.16 & \multicolumn{1}{c|}{20.95} & 19.75 \\
      JPEG & \multicolumn{1}{c|}{23.21} & 22.70 & \multicolumn{1}{c|}{21.99} & 21.59 & \multicolumn{1}{c|}{21.91} & 21.84 & \multicolumn{1}{c|}{18.75} & 19.24 & \multicolumn{1}{c|}{20.51} & 19.92 \\ \hline 
      \end{tabular}
      }
  \end{center}
  \vskip -0.3cm 
  \label{table:oneClient}
  \vskip -0.30cm
\end{table*}

\subsection*{Tables for Varying Number of Clients}
This section presents the quantitative results on varying numbers of clients.
\autoref{table:degradations_srresnet} shows the PSNR results for SRResNet and \autoref{table:degradations_rrdb} shows the PSNR results for RRDB.

\begin{table*}[ht!]
  \caption{PSNR results of SRResNet with $\times 4$ scaling and different number of clients.
  }
  \vskip -0.7cm 
  \small
  \begin{center}
  \resizebox{.90\textwidth}{!}{%
    \begin{tabular}{|c|cc|cc|cc|cc|cc|}
      \hline
       & \multicolumn{2}{c|}{Set5~\cite{Set5}} & \multicolumn{2}{c|}{Set14~\cite{Set14}} & \multicolumn{2}{c|}{BSD100~\cite{BSD100}} & \multicolumn{2}{c|}{Manga109~\cite{Manga109}} & \multicolumn{2}{c|}{Urban100~\cite{Urban100}} \\ \hline
      Clients & \multicolumn{1}{c|}{clean} & blur & \multicolumn{1}{c|}{clean} & blur & \multicolumn{1}{c|}{clean} & blur & \multicolumn{1}{c|}{clean} & blur & \multicolumn{1}{c|}{clean} & blur \\ \hline
      
      4 & \multicolumn{1}{c|}{28.59} & 25.80 & \multicolumn{1}{c|}{25.59} & 23.74 & \multicolumn{1}{c|}{24.65} & 23.91 & \multicolumn{1}{c|}{19.43} & 20.42 & \multicolumn{1}{c|}{23.47} & 21.54 \\
      8 & \multicolumn{1}{c|}{28.55} & \textbf{26.05} & \multicolumn{1}{c|}{25.54} & \textbf{23.94} & \multicolumn{1}{c|}{\textbf{24.67}} & \textbf{24.01} & \multicolumn{1}{c|}{\textbf{19.43}} & \textbf{20.42} & \multicolumn{1}{c|}{23.39} & \textbf{21.69} \\
      12 & \multicolumn{1}{c|}{28.44} & 26.04 & \multicolumn{1}{c|}{25.50} & 23.93 & \multicolumn{1}{c|}{\textbf{24.69}} & \textbf{24.01} & \multicolumn{1}{c|}{\textbf{19.53}} & \textbf{20.43} & \multicolumn{1}{c|}{23.31} & 21.68 \\
      16 & \multicolumn{1}{c|}{28.42} & 25.96 & \multicolumn{1}{c|}{25.49} & 23.87 & \multicolumn{1}{c|}{\textbf{24.70}} & 23.99 & \multicolumn{1}{c|}{\textbf{19.65}} & 20.39 & \multicolumn{1}{c|}{23.27} & 21.60 \\ \hline \hline
      
      Clients & \multicolumn{1}{c|}{noise} & jpeg & \multicolumn{1}{c|}{noise} & jpeg & \multicolumn{1}{c|}{noise} & jpeg & \multicolumn{1}{c|}{noise} & jpeg & \multicolumn{1}{c|}{noise} & jpeg \\ \hline
  
      4 & \multicolumn{1}{c|}{20.18} & 24.31 & \multicolumn{1}{c|}{19.43} & 22.90 & \multicolumn{1}{c|}{19.40} & 23.10 & \multicolumn{1}{c|}{17.53} & 18.97 & \multicolumn{1}{c|}{18.48} & 21.23 \\
      8 & \multicolumn{1}{c|}{\textbf{20.91}} & \textbf{24.34} & \multicolumn{1}{c|}{\textbf{20.15}} & \textbf{22.90} & \multicolumn{1}{c|}{\textbf{20.24}} & \textbf{23.11} & \multicolumn{1}{c|}{\textbf{17.99}} & \textbf{19.00} & \multicolumn{1}{c|}{\textbf{19.13}} & \textbf{21.24} \\
      12 & \multicolumn{1}{c|}{\textbf{21.17}} & \textbf{24.38} & \multicolumn{1}{c|}{\textbf{20.34}} & \textbf{22.94} & \multicolumn{1}{c|}{\textbf{20.37}} & \textbf{23.14} & \multicolumn{1}{c|}{\textbf{18.09}} & \textbf{19.08} & \multicolumn{1}{c|}{\textbf{19.26}} & \textbf{21.28} \\
      16 & \multicolumn{1}{c|}{\textbf{21.21}} & \textbf{24.43} & \multicolumn{1}{c|}{\textbf{20.34}} & \textbf{22.98} & \multicolumn{1}{c|}{20.31} & \textbf{23.17} & \multicolumn{1}{c|}{\textbf{18.11}} & \textbf{19.16} & \multicolumn{1}{c|}{19.23} & \textbf{21.31} \\ \hline \hline
      
      Clients & \multicolumn{1}{c|}{b+n} & b+j & \multicolumn{1}{c|}{b+n} & b+j & \multicolumn{1}{c|}{b+n} & b+j & \multicolumn{1}{c|}{b+n} & b+j & \multicolumn{1}{c|}{b+n} & b+j \\ \hline
      
      4 & \multicolumn{1}{c|}{23.15} & 23.46 & \multicolumn{1}{c|}{22.00} & 22.24 & \multicolumn{1}{c|}{22.12} & 22.80 & \multicolumn{1}{c|}{19.51} & 19.47 & \multicolumn{1}{c|}{20.30} & 20.35 \\
      8 & \multicolumn{1}{c|}{\textbf{23.48}} & \textbf{23.46} & \multicolumn{1}{c|}{\textbf{22.28}} & 22.23 & \multicolumn{1}{c|}{\textbf{22.49}} & \textbf{22.80} & \multicolumn{1}{c|}{\textbf{19.67}} & \textbf{19.47} & \multicolumn{1}{c|}{\textbf{20.53}} & \textbf{20.36} \\
      12 & \multicolumn{1}{c|}{\textbf{23.64}} & \textbf{23.50} & \multicolumn{1}{c|}{\textbf{22.36}} & \textbf{22.27} & \multicolumn{1}{c|}{\textbf{22.55}} & \textbf{22.81} & \multicolumn{1}{c|}{\textbf{19.73}} & \textbf{19.50} & \multicolumn{1}{c|}{\textbf{20.58}} & \textbf{20.39} \\
      16 & \multicolumn{1}{c|}{\textbf{23.66}} & 23.48 & \multicolumn{1}{c|}{\textbf{22.38}} & 22.25 & \multicolumn{1}{c|}{\textbf{22.58}} & \textbf{22.81} & \multicolumn{1}{c|}{\textbf{19.74}} & 19.49 & \multicolumn{1}{c|}{\textbf{20.58}} & 20.37 \\ \hline \hline
      
      Clients & \multicolumn{1}{c|}{n+j} & b+n+j & \multicolumn{1}{c|}{n+j} & b+n+j & \multicolumn{1}{c|}{n+j} & b+n+j & \multicolumn{1}{c|}{n+j} & b+n+j & \multicolumn{1}{c|}{n+j} & b+n+j \\ \hline
      
      4 & \multicolumn{1}{c|}{22.12} & 22.04 & \multicolumn{1}{c|}{21.18} & 21.04 & \multicolumn{1}{c|}{21.24} & 21.40 & \multicolumn{1}{c|}{18.20} & 18.80 & \multicolumn{1}{c|}{19.79} & 19.52 \\
      8  & \multicolumn{1}{c|}{\textbf{22.28}} & \textbf{22.08} & \multicolumn{1}{c|}{\textbf{21.28}} & \textbf{21.09} & \multicolumn{1}{c|}{\textbf{21.36}} & \textbf{21.45} & \multicolumn{1}{c|}{\textbf{18.31}} & \textbf{18.84} & \multicolumn{1}{c|}{\textbf{19.90}} & \textbf{19.55} \\
      12  & \multicolumn{1}{c|}{\textbf{22.38}} & \textbf{22.12} & \multicolumn{1}{c|}{\textbf{21.34}} & \textbf{21.12} & \multicolumn{1}{c|}{\textbf{21.39}} & \textbf{21.45} & \multicolumn{1}{c|}{\textbf{18.40}} & \textbf{18.87} & \multicolumn{1}{c|}{\textbf{19.96}} & \textbf{19.58} \\
      16  & \multicolumn{1}{c|}{\textbf{22.54}} & \textbf{22.19} & \multicolumn{1}{c|}{\textbf{21.49}} & 21.10 & \multicolumn{1}{c|}{\textbf{21.52}} & \textbf{21.52} & \multicolumn{1}{c|}{\textbf{18.52}} & \textbf{18.91} & \multicolumn{1}{c|}{\textbf{20.08}} & \textbf{19.62} \\ \hline 
      \end{tabular}
      }
  \end{center}
  \label{table:degradations_srresnet}
  \vskip -0.50cm
\end{table*}

\begin{table*}[ht!]
  \caption{PSNR results of RRDB with $\times 4$ scaling and a varying number of clients. 
  }
  \vskip -0.7cm 
  \small
  \begin{center}
  \resizebox{.90\textwidth}{!}{%
    \begin{tabular}{|c|cc|cc|cc|cc|cc|}
      \hline
       & \multicolumn{2}{c|}{Set5~\cite{Set5}} & \multicolumn{2}{c|}{Set14~\cite{Set14}} & \multicolumn{2}{c|}{BSD100~\cite{BSD100}} & \multicolumn{2}{c|}{Manga109~\cite{Manga109}} & \multicolumn{2}{c|}{Urban100~\cite{Urban100}} \\ \hline
      Clients & \multicolumn{1}{c|}{clean} & blur & \multicolumn{1}{c|}{clean} & blur & \multicolumn{1}{c|}{clean} & blur & \multicolumn{1}{c|}{clean} & blur & \multicolumn{1}{c|}{clean} & blur \\ \hline
      
      4 & \multicolumn{1}{c|}{28.53} & 26.02 & \multicolumn{1}{c|}{25.50} & 23.90 & \multicolumn{1}{c|}{24.49} & 23.99 & \multicolumn{1}{c|}{18.67} & 20.48 & \multicolumn{1}{c|}{23.50} & 21.87 \\
      8 & \multicolumn{1}{c|}{\textbf{28.72}} & \textbf{26.07} & \multicolumn{1}{c|}{\textbf{25.70}} & \textbf{23.92} & \multicolumn{1}{c|}{\textbf{24.63}} & \textbf{24.01} & \multicolumn{1}{c|}{\textbf{19.00}} & \textbf{20.49} & \multicolumn{1}{c|}{\textbf{23.67}} & 21.75 \\
      12 & \multicolumn{1}{c|}{\textbf{28.75}} & \textbf{26.13} & \multicolumn{1}{c|}{25.68} & \textbf{23.95} & \multicolumn{1}{c|}{\textbf{24.66}} & \textbf{24.04} & \multicolumn{1}{c|}{\textbf{19.07}} & 20.47 & \multicolumn{1}{c|}{23.62} & \textbf{21.79} \\
      16 & \multicolumn{1}{c|}{\textbf{28.78}} & \textbf{26.13} & \multicolumn{1}{c|}{\textbf{25.69}} & \textbf{23.97} & \multicolumn{1}{c|}{\textbf{24.66}} & 24.02 & \multicolumn{1}{c|}{\textbf{19.14}} & 20.43 & \multicolumn{1}{c|}{23.61} & 21.77 \\ \hline \hline
      
      Clients & \multicolumn{1}{c|}{noise} & jpeg & \multicolumn{1}{c|}{noise} & jpeg & \multicolumn{1}{c|}{noise} & jpeg & \multicolumn{1}{c|}{noise} & jpeg & \multicolumn{1}{c|}{noise} & jpeg \\ \hline
  
      4 & \multicolumn{1}{c|}{19.94} & 24.04 & \multicolumn{1}{c|}{19.37} & 22.68 & \multicolumn{1}{c|}{19.45} & 22.97 & \multicolumn{1}{c|}{17.35} & 18.61 & \multicolumn{1}{c|}{18.57} & 20.94 \\
      8  & \multicolumn{1}{c|}{\textbf{20.53}} & \textbf{24.13} & \multicolumn{1}{c|}{\textbf{19.83}} & \textbf{22.75} & \multicolumn{1}{c|}{\textbf{19.92}} & \textbf{23.00} & \multicolumn{1}{c|}{\textbf{17.65}} & \textbf{18.76} & \multicolumn{1}{c|}{\textbf{18.98}} & \textbf{21.06} \\
      12  & \multicolumn{1}{c|}{20.43} & \textbf{24.22} & \multicolumn{1}{c|}{19.73} & \textbf{22.81} & \multicolumn{1}{c|}{19.81} & \textbf{23.04} & \multicolumn{1}{c|}{\textbf{17.65}} &  \textbf{18.84} & \multicolumn{1}{c|}{18.85} & \textbf{21.14} \\
      16  & \multicolumn{1}{c|}{\textbf{20.55}} & \textbf{24.30} & \multicolumn{1}{c|}{\textbf{19.77}} & \textbf{22.89} & \multicolumn{1}{c|}{19.76} & \textbf{23.08} & \multicolumn{1}{c|}{\textbf{17.68}} & \textbf{18.90} & \multicolumn{1}{c|}{\textbf{18.87}} & \textbf{21.21} \\ \hline \hline
      
      Clients & \multicolumn{1}{c|}{b+n} & b+j & \multicolumn{1}{c|}{b+n} & b+j & \multicolumn{1}{c|}{b+n} & b+j & \multicolumn{1}{c|}{b+n} & b+j & \multicolumn{1}{c|}{b+n} & b+j \\ \hline
      
      4 & \multicolumn{1}{c|}{22.95} & 23.44 & \multicolumn{1}{c|}{21.86} & 22.22 & \multicolumn{1}{c|}{21.97} & 22.80 & \multicolumn{1}{c|}{19.40} & 19.45 & \multicolumn{1}{c|}{20.37} & 20.36 \\
      8 & \multicolumn{1}{c|}{\textbf{23.34}} & \textbf{23.44} & \multicolumn{1}{c|}{\textbf{22.15}} & \textbf{22.22} & \multicolumn{1}{c|}{\textbf{22.26}} & \textbf{22.80} & \multicolumn{1}{c|}{\textbf{19.62}} & \textbf{19.46} & \multicolumn{1}{c|}{\textbf{20.56}} & \textbf{20.36} \\
      12 & \multicolumn{1}{c|}{23.23} & \textbf{23.47} & \multicolumn{1}{c|}{22.14} & \textbf{22.24} & \multicolumn{1}{c|}{22.23} & \textbf{22.81} & \multicolumn{1}{c|}{19.61} &  \textbf{19.48} & \multicolumn{1}{c|}{20.54} & \textbf{20.38} \\
      16 & \multicolumn{1}{c|}{\textbf{23.39}} & \textbf{23.47} & \multicolumn{1}{c|}{\textbf{22.18}} & \textbf{22.25} & \multicolumn{1}{c|}{\textbf{22.27}} & \textbf{22.81} & \multicolumn{1}{c|}{\textbf{19.63}} & \textbf{19.48} & \multicolumn{1}{c|}{20.52} & \textbf{20.38}  \\ \hline \hline
      
      Clients & \multicolumn{1}{c|}{n+j} & b+n+j & \multicolumn{1}{c|}{n+j} & b+n+j & \multicolumn{1}{c|}{n+j} & b+n+j & \multicolumn{1}{c|}{n+j} & b+n+j & \multicolumn{1}{c|}{n+j} & b+n+j \\ \hline

      4 & \multicolumn{1}{c|}{21.47} & 21.82 & \multicolumn{1}{c|}{20.70} & 20.87 & \multicolumn{1}{c|}{20.84} & 21.22 & \multicolumn{1}{c|}{17.74} & 18.68 & \multicolumn{1}{c|}{19.26} & 19.38 \\
      8 & \multicolumn{1}{c|}{\textbf{21.70}} & \textbf{21.87} & \multicolumn{1}{c|}{\textbf{20.88}} & \textbf{20.91} & \multicolumn{1}{c|}{\textbf{20.97}} & \textbf{21.26} & \multicolumn{1}{c|}{\textbf{17.95}} & \textbf{18.73} & \multicolumn{1}{c|}{\textbf{19.49}} & \textbf{19.43} \\
      12 & \multicolumn{1}{c|}{\textbf{21.89}} & \textbf{21.92} & \multicolumn{1}{c|}{\textbf{20.99}} & \textbf{20.95} & \multicolumn{1}{c|}{\textbf{21.05}} & \textbf{21.28} & \multicolumn{1}{c|}{\textbf{18.06}} & \textbf{18.77} & \multicolumn{1}{c|}{\textbf{19.61}} &  \textbf{19.45} \\
      16 & \multicolumn{1}{c|}{\textbf{22.08}} & \textbf{22.00} & \multicolumn{1}{c|}{\textbf{21.14}} & \textbf{21.03} & \multicolumn{1}{c|}{\textbf{21.18}} & \textbf{21.36} & \multicolumn{1}{c|}{\textbf{18.17}} & \textbf{18.81} & \multicolumn{1}{c|}{\textbf{19.76}} &  \textbf{19.51} \\ \hline 
      \end{tabular}
  }
  \end{center}
  \label{table:degradations_rrdb}
  \vskip -0.50cm
\end{table*}

\newpage

\subsection*{Additional Visualizations and Results on Different Degradation Distributions (relative to uniform distribution)}

\begin{figure}[ht!]
  \centering
  \includegraphics[width=\linewidth]{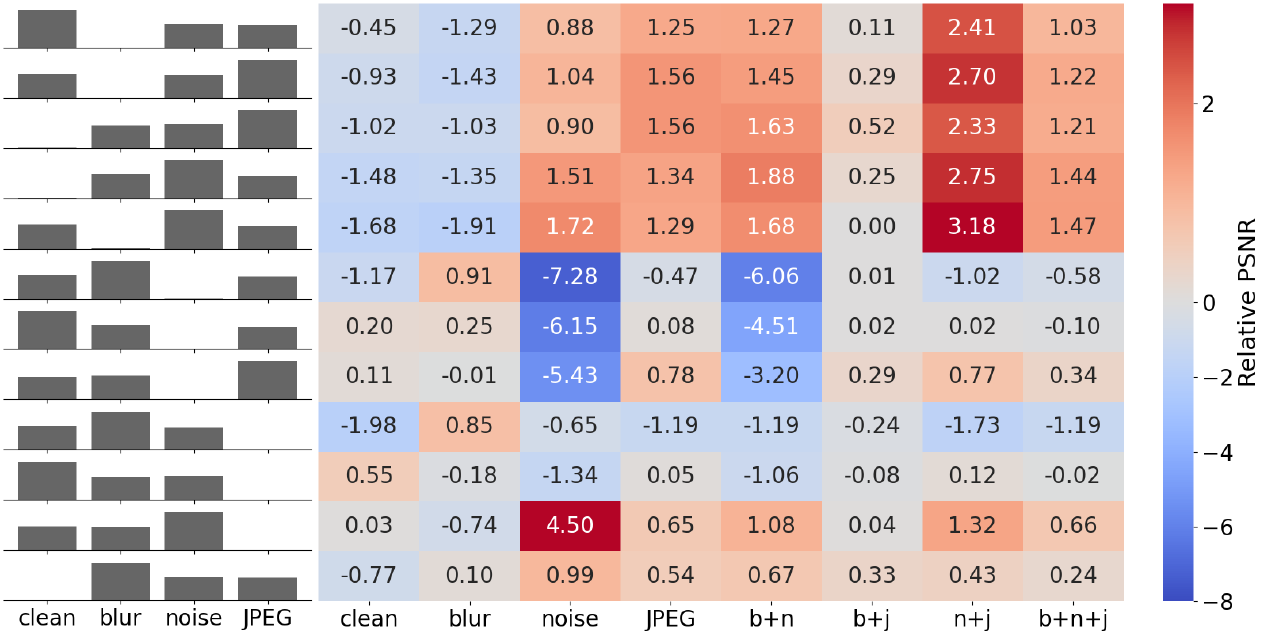}
  \caption{\label{fig:dataDistribution_set5_rel}Experimental results of different degradation distributions on Set5.  
  On the left side is the degradation distribution during training.
  The resulting performance on the respective degraded test set is on the right side (relative to uniform distribution).
}
\end{figure}

\begin{figure}[ht!]
  \centering
  \includegraphics[width=\linewidth]{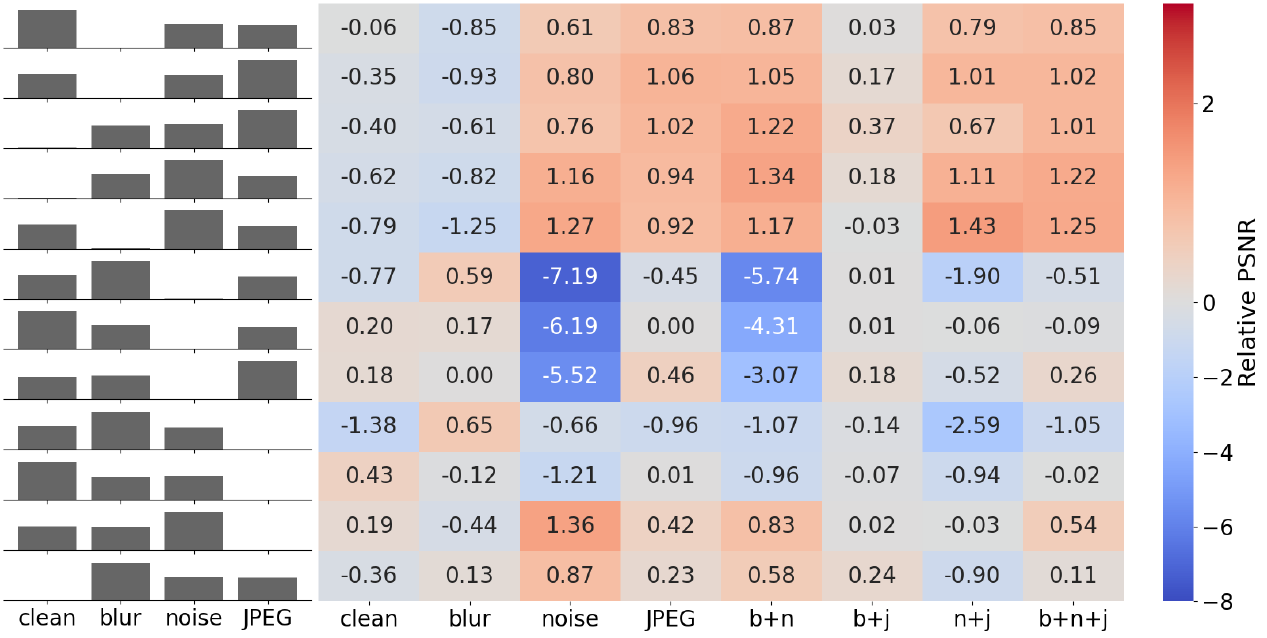}
  \caption{\label{fig:dataDistribution_set14_rel}Experimental results of different degradation distributions on Set14.  
  On the left side is the degradation distribution during training.
  The resulting performance on the respective degraded test set is on the right side (relative to uniform distribution).
}
\end{figure}

\begin{figure}[ht!]
  \centering
  \includegraphics[width=\linewidth]{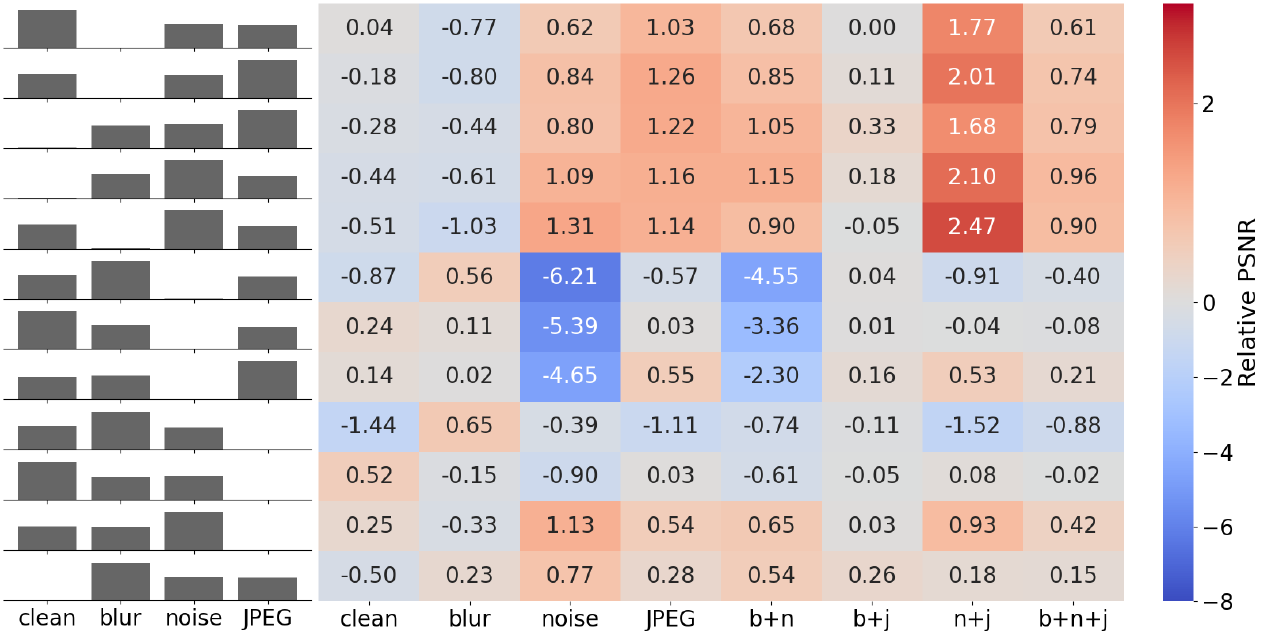}
  \caption{\label{fig:dataDistribution_urbann_rel}Experimental results of different degradation distributions on Urban100.  
  On the left side is the degradation distribution during training.
  The resulting performance on the respective degraded test set is on the right side (relative to uniform distribution).
}
\end{figure}

\begin{figure}[ht!]
  \centering
  \includegraphics[width=\linewidth]{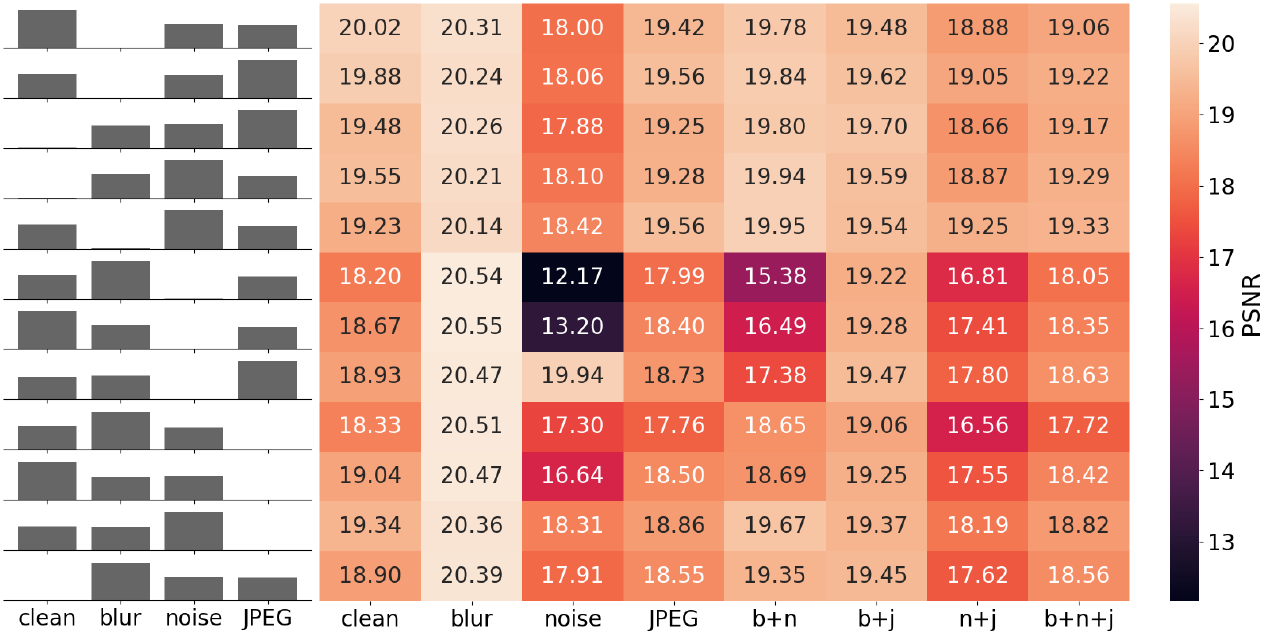}
  \caption{\label{fig:dataDistribution_manga109_rel}Experimental results of different degradation distributions on Manga109.  
  On the left side is the degradation distribution during training.
  The resulting performance on the respective degraded test set is on the right side (relative to uniform distribution).
}
\end{figure}

\clearpage

\subsection*{Additional Visualizations and Results on Different Degradation Distributions}

\begin{figure}[ht!]
  \centering
  \includegraphics[width=\linewidth]{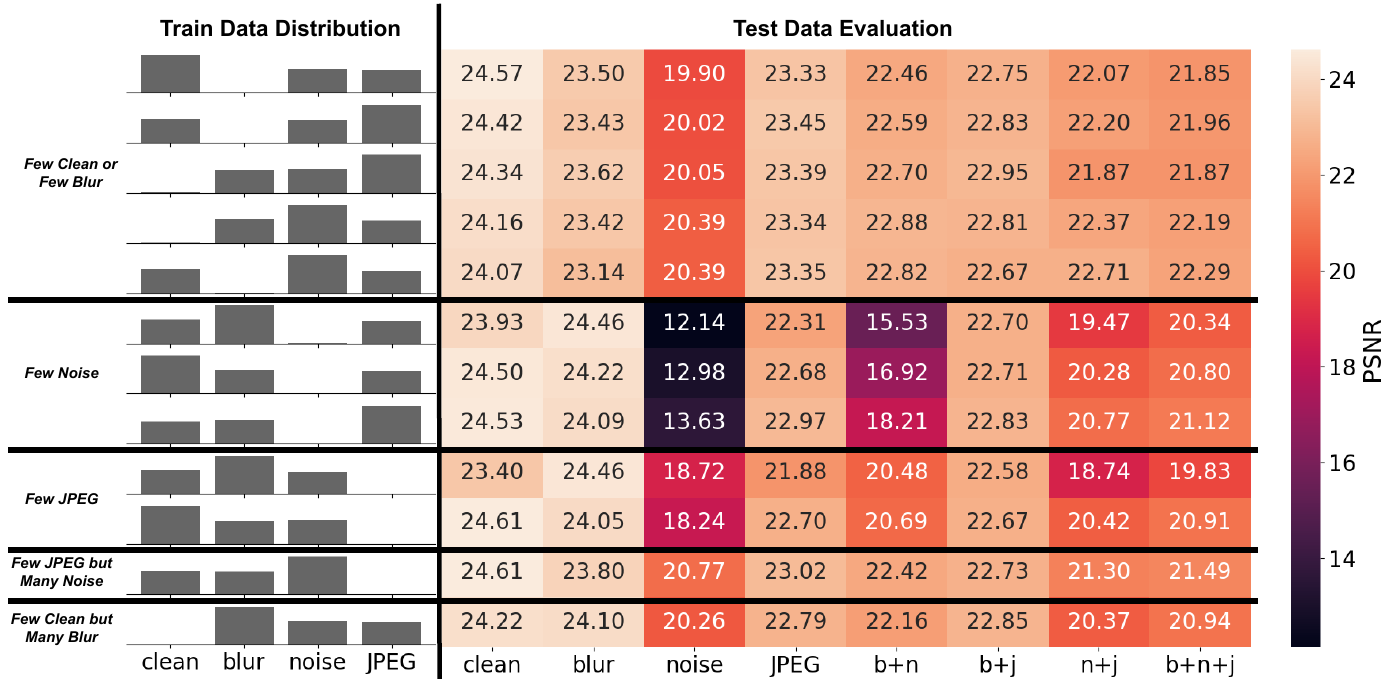}
  \caption{\label{fig:dataDistribution_BSD}Experimental results of different degradation distributions on BSD100.  
  On the left side is the degradation distribution during training.
  The resulting performance on the respective degraded test set is on the right side.
}
\end{figure}

\begin{figure}[ht!]
  \centering
  \includegraphics[width=\linewidth]{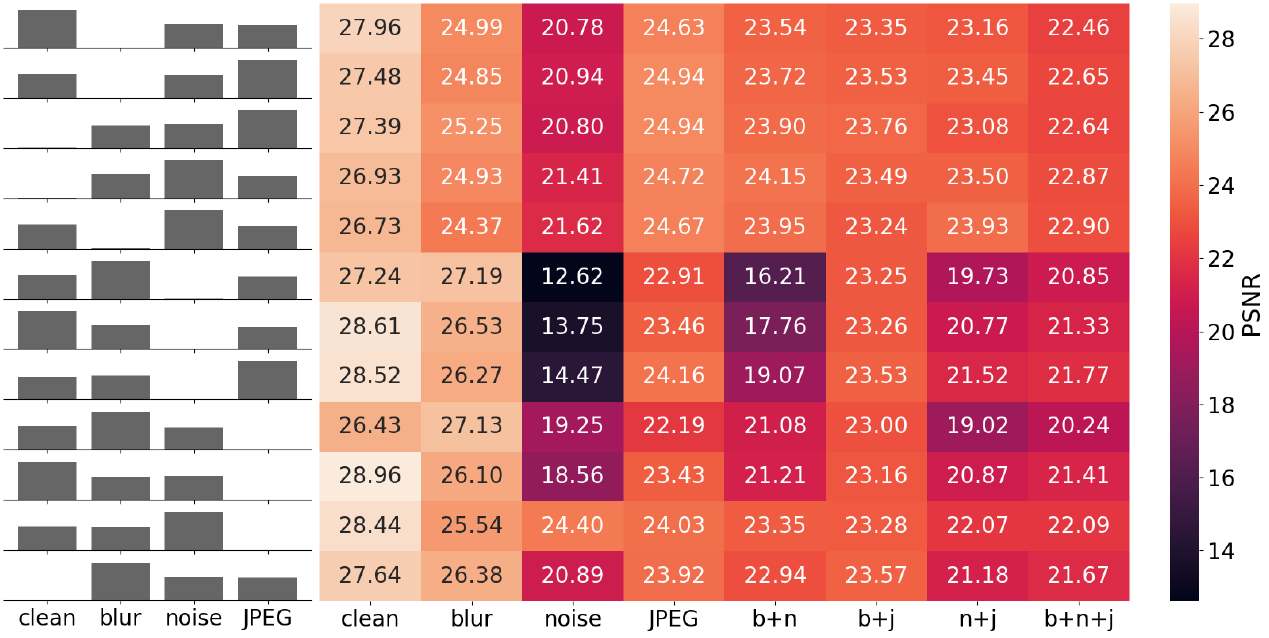}
  \caption{\label{fig:dataDistribution_set5}Experimental results of different degradation distributions on Set5.  
  On the left side is the degradation distribution during training.
  The resulting performance on the respective degraded test set is on the right side.
}
\end{figure}

\begin{figure}[ht!]
  \centering
  \includegraphics[width=\linewidth]{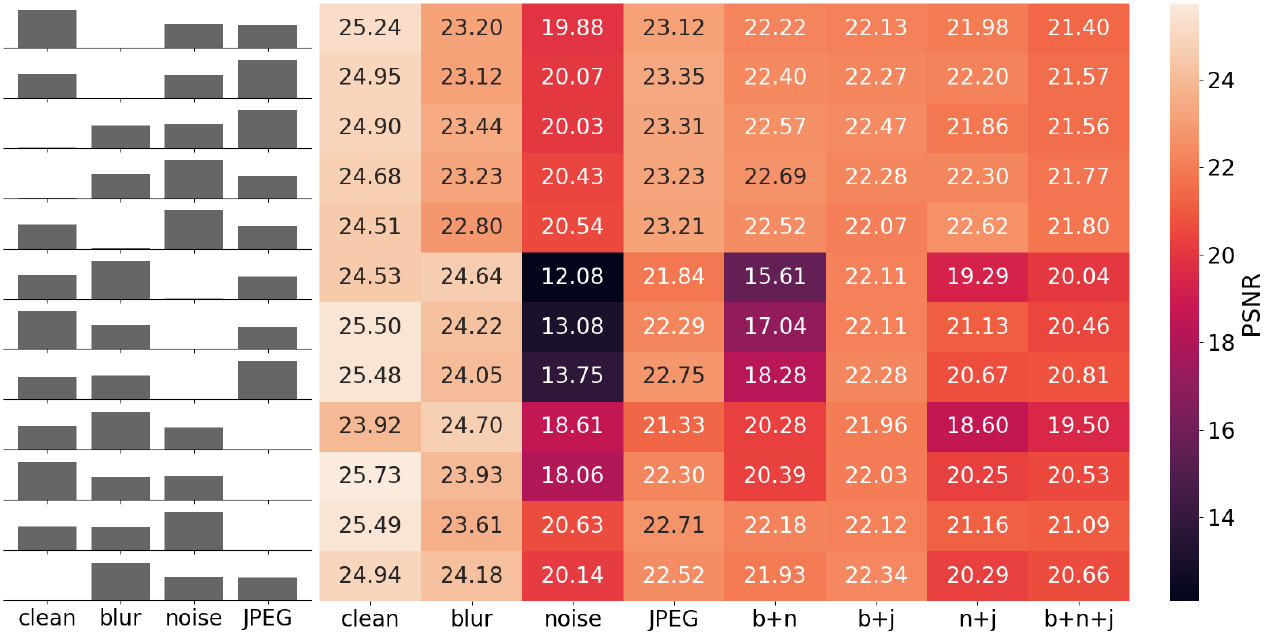}
  \caption{\label{fig:dataDistribution_set14}Experimental results of different degradation distributions on Set14.  
  On the left side is the degradation distribution during training.
  The resulting performance on the respective degraded test set is on the right side.
}
\end{figure}

\begin{figure}[ht!]
  \centering
  \includegraphics[width=\linewidth]{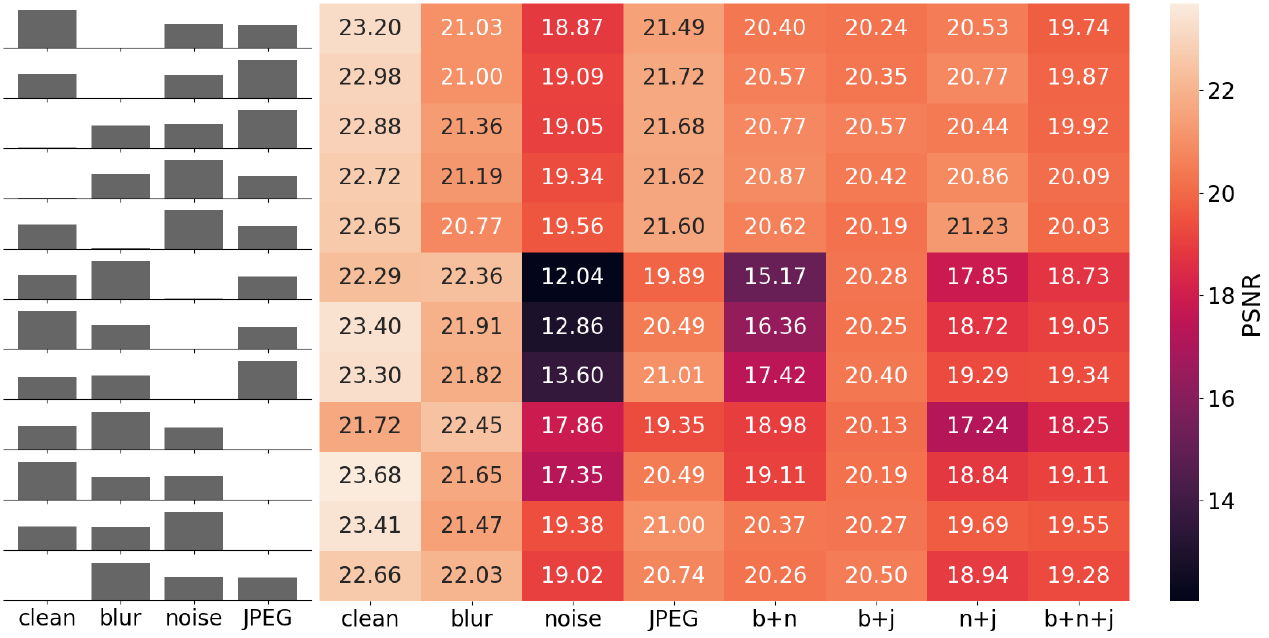}
  \caption{\label{fig:dataDistribution_urban100}Experimental results of different degradation distributions on Urban100.  
  On the left side is the degradation distribution during training.
  The resulting performance on the respective degraded test set is on the right side.
}
\end{figure}

\clearpage

\begin{figure}[ht!]
  \centering
  \includegraphics[width=\linewidth]{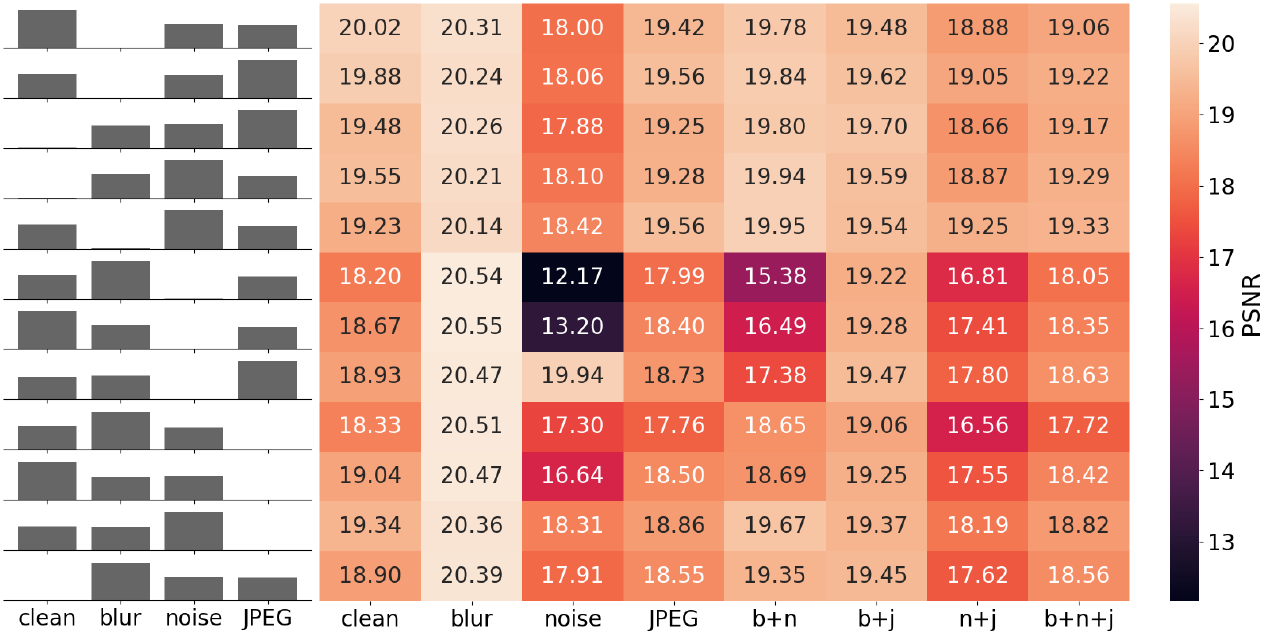}
  \caption{\label{fig:dataDistribution_manga109}Experimental results of different degradation distributions on Manga109.  
  On the left side is the degradation distribution during training.
  The resulting performance on the respective degraded test set is on the right side.
}
\end{figure}

\end{document}